\documentclass[]{aa}  

\usepackage{graphicx}
\usepackage{txfonts}
\usepackage{amsmath}
\usepackage{siunitx}
\usepackage{longtable}
\usepackage{color}
\usepackage[bottom]{footmisc}
                                 
\newcommand{\ca}[1]{{#1}}

\usepackage{hyperref}

\hypersetup{
   bookmarks=true,         
   unicode=true,           
   pdftoolbar=true,        
   pdfmenubar=true,        
   pdffitwindow=true,      
   pdftitle={TITLE },
   pdfauthor={YOUR NAME},
   pdfsubject={Planetary Science},
   pdfkeywords={},         
   pdfnewwindow=true,      
   colorlinks=true,        
   linkcolor=blue,         
   citecolor=blue,         
   filecolor=green,         
   urlcolor=green          
}

\begin{document}

   \title{An ancient and a primordial collisional family as the main sources of X-type asteroids of the inner Main Belt
   \thanks{This work is dedicated to the memory of Andrea Milani, 
                who put the foundation and devoted his scientific career on the study of asteroid families.}}


   \author{Marco Delbo'
          \inst{1}
          \and
          Chrysa Avdellidou
          \inst{1,2}
          \and  Alessandro Morbidelli  \inst{1}
          }

   \institute{Universit\'e C\^ote d'Azur, CNRS--Lagrange, Observatoire de la C\^ote d'Azur,
              CS 34229 -- F 06304 NICE Cedex 4, France\\
              \email{marcodelbo@gmail.com}
              \email{chavdell@gmail.com}
         \and
             Science Support Office, Directorate of Science, European Space Agency, 
             Keplerlaan 1, NL-2201 AZ Noordwijk ZH, The Netherlands.\\
             }

   \date{Received \today; accepted \today}

   \titlerunning{X-type families in the inner Main Belt}
   
 
  \abstract
   {}
   {The near-Earth asteroid population suggests the existence of an inner Main Belt source of asteroids that belongs to the spectroscopic X-complex and has moderate albedos. The identification of such a source has been lacking so far. We argue that the most probable source is one or more collisional asteroid families that escaped discovery up to now.}
   {We apply a novel method to search for asteroid families in the inner Main Belt population of asteroids belonging to the X-complex with moderate albedo. Instead of searching for asteroid clusters in orbital elements space, which could be severely dispersed when older than some billions of years, our method looks for correlations between the orbital semimajor axis and the inverse size of asteroids. This correlation is the signature of members of collisional families, which drifted from a common centre under the effect of the Yarkovsky thermal effect.}
   {We identify two previously unknown families in the inner Main Belt among the moderate-albedo X-complex asteroids. One of them, \ca{whose lowest numbered asteroid is (161) Athor,} is $\sim$3~Gyrs-old, whereas the second one, \ca{whose lowest numbered object is (689) Zita,} can be as old as the Solar System. Members of this latter family have orbital eccentricities and inclinations that spread them over the entire inner Main Belt, which is an indication that this family could be primordial, i.e. it formed before the giant planet orbital instability.}
   {The vast majority of moderate-albedo X-complex asteroids of the inner-Main Belt are genetically related, as they can be included into few asteroid families. 
   Only nine X-complex asteroids with moderate albedo of the inner Main Belt cannot be included in asteroid families. We suggest that these bodies formed by direct accretion of the solids in the protoplanetary disk, and are thus surviving planetesimals.}


   \maketitle
%

\section{Introduction\label{S:intro}}

Collisions in the asteroid Main Belt are responsible for sculpting the size distribution of these bodies \citep{Bottke2015aste.book..701B},  forming craters on their surfaces, producing fresh regolith \citep{Horz1997M&PS...32..179H,Basilevsky2015P&SS..117..312B}, ejecting asteroid material into space \citep{Jewitt2011ApJ...733L...4J}, and also implanting exogenous materials \citep{McCord2012Natur.491...83M,Avdellidou2018MNRAS.475.3419A,Avdellidou2017MNRAS.464..734A,Avdellidou2016MNRAS.456.2957A,turrini2016, Vernazza2017AJ....153...72V}. 

The most energetic impacts can eject asteroid fragments at speeds larger than the gravitational escape velocity of the parent body. This process can form families of daughter asteroids initially placed on orbits that group near that of the parent asteroid \citep{zappala1984}. Asteroid families can therefore be recognised as clusters of bodies in proper orbital element space -- proper semimajor axis, proper eccentricity, and proper inclination $(a,e,i)$ -- with significant contrast with respect to the local background \citep{Milani2014Icar..239...46M,Nesvorny2015aste.book..297N}. The Hierarchical Clustering Method \citep[HCM;][and references therein]{Zappala1990AJ....100.2030Z,Nesvorny2015aste.book..297N} is typically used for the identification of these asteroid clusters. 

However, family members disperse with time. Dispersion is driven by a change in the $a$-value (d$a$/dt $\neq$ 0) due to the thermal radiation force of the Yarkovsky effect \citep{Vokrouhlicky2006Icar..182..118V}; the eccentricity and inclination are affected by orbital resonances with the planets, the locations of which are crossed by the asteroids as they drift in semimajor axis. This process has  fundamental consequences: 
Families older than $\sim$2 Gyr tend to lose \ca{number density} contrast with respect to the local background population \citep{Parker2008Icar..198..138P, Spoto2015Icar..257..275S, Carruba2016MNRAS.458.3731C}. They become more difficult to be detected by the HCM compared to younger ones \citep{Walsh2013Icar..225..283W,Bolin2017Icar..282..290B,Delbo2017Sci...357.1026M}. In addition to the effect of orbital resonances currently present in the Main Belt, the scattering of fragments of primordial families \citep{MIlani2017Icar..288..240M,Milani2014Icar..239...46M,Delbo2017Sci...357.1026M} was also affected by major dynamical events such as the giant planet orbital instability, for which we have evidence that happened at some point in the solar system history \citep{Morbidelli2015aste.book..493M}. Said instability shifted the positions of the resonances and resulted in the incoherent dispersion of $e$ and $i$ of any pre-existing primordial asteroid family in the Main Belt \citep{OBrasil2016Icar..266..142B}, as it is the case for family of low-albedo asteroids found by \cite{Delbo2017Sci...357.1026M} in the inner-Main Belt (i.e. 2.1 $< a <$ 2.5 au).

The sign of d$a$/d$t$ depends on the obliquity of the asteroid's spin vector, with prograde-rotating asteroids drifting with d$a$/d$t>0$ and retrograde ones with d$a$/d$t<0$. To a first order, the value of d$a$/d$t$ is inversely proportional to the diameter $D$ of family members, such that, at any given epoch, the orbits of smaller asteroids are moved further away from the family's centre than the bigger ones. Due to this mechanism, an asteroid family forms a characteristic shape in the space of proper semimajor axis $a$ vs. inverse diameter $(a,1/D)$, which is called ``V-shape'' because the distribution of asteroids resemble the letter ``V'' \citep{Milani2014Icar..239...46M, Spoto2015Icar..257..275S,Bolin2017Icar..282..290B}. The slopes of the borders of the ``V'' indicate the age of a family \citep[see e.g.][]{Spoto2015Icar..257..275S}, with younger ones having steep and older ones shallow slopes. While orbital resonance crossings produce diffusion of $e$ and $i$ these \ca{have minimal effect on $a$ \citep{MilicZitnik2016ApJ...816L..31M}}, resulting in the conservation of the V-shape of families for billions of years. 
The semimajor axis values can only be modified by gravitational scattering due to close encounters with massive asteroids \citep{Carruba2013MNRAS.433.2075C,Delisle2012A&A...540A.118D} or in the case that a planet momentarily entered the Main Belt during the orbital instability phases of the giant planets \citep{OBrasil2016Icar..266..142B}. The former effect has been shown to be negligible for $\sim$Gyr-old families \citep{Delbo2017Sci...357.1026M}. The second case did not happen for the inner Main Belt, otherwise the V-shape of the primordial family discovered by \cite{Delbo2017Sci...357.1026M} would not be visible. 

Another fundamental consequence of the $a$-mobility due to the Yarkovsky effect is that asteroids can drift into powerful mean motion or secular orbital resonances with the planets, causing these bodies to leave the Main Belt. These escaping asteroids can reach orbits in the inner Solar System, eventually becoming near-Earth asteroids (NEAs) \citep{Morbidelli2003Icar..163..120M}. Ideally, one would like to trace back the orbital evolution of NEAs and find their place of origin in the Main Belt. Unfortunately, this is not possible in a deterministic way due to the chaotic nature of their orbital evolution. Still not all the information is lost, and methods have been developed to identify the source region of NEAs in a statistical sense \citep{Bottke2002Icar..156..399B,Granvik2017A&A...598A..52G,Granvik2016Natur.530..303G,Greenstreet2012Icar..217..355G}. For each NEA, a probability to originate from different source regions, which include the Main Belt, the Jupiter Family Comets (JFC), the Hungaria (HU) and the Phocaea (PHO) populations are calculated. It is found that the most efficient route from the Main Belt to near-Earth space is offered by the family formation in the inner Main Belt and delivery through the $\nu_6$ resonance complex \citep{Granvik2016Natur.530..303G} or -- to lesser extent -- the J3:1 mean motion resonance (MMR) with Jupiter. 

On the basis of the source region probabilities, osculating orbital elements, spectral classes, and albedos, different studies have attempted to 
match properties of notable NEAs with their main belt family counterparts. For instance, the NEA (3200) Phaethon, which is associated with the Geminids meteor stream \citep{Fox1984MNRAS.208P..11F,Gustafson1989A&A...225..533G,Williams1993MNRAS.262..231W}, has been linked to the Pallas family \citep{DeLeon2010A&A...513A..26D,Todorovic2018MNRAS.475..601T}; (101955) Bennu, target of NASA's OSIRIS-REx sample return mission \citep{Lauretta2012LPICo1667.6291L}, is likely coming from one of the low-albedo, low-orbital inclination families of the inner Main Belt such as Eulalia or Polana \citep{Bottke2015Icar..247..191B,Campins2010ApJ...721L..53C}. The target of JAXA's Hayabusa2 sample return mission (162173) Ryugu, also identified as 1999 JU$_3$, was linked to the Polana family \citep{Campins2013AJ....146...26C} or the low-albedo asteroid background of the inner Main Belt. The latter has been later suggested to form a family by itself that can be as old as the Solar System \citep{Delbo2017Sci...357.1026M}. The discovery of the aforementioned primordial family implies that the background of low-albedo unaffiliated asteroids in the inner Main Belt is represented only by few asteroids which are all larger than $\sim$50~km in diameter. This means that the smaller low-albedo asteroids of the inner Main Belt could be genetically linked to a few distinct asteroid parents \citep{Delbo2017Sci...357.1026M} and that the low-albedo NEAs with high probability to come from the inner Main Belt are also linked to these few distinct asteroid parents. These later findings are also independently confirmed by the work of \cite{Dermott2018NatAs...2..549D}.

On the other hand, concerning the NEAs with high albedo ($p_V>0.12$) and not belonging to the spectroscopic C-complex, the situation is less clear. Nonetheless, several important S-complex families in the inner Main Belt are known that could be the source of NEAs belonging to the same spectroscopic complex. For instance, the Flora family \citep{Vernazza2008Natur.454..858V}  is capable of delivering NEAs belonging to the S-complex and with composition similar to the LL ordinary chondrite meteorites \citep{Vokrouhlicky2017AJ....153..172V}. Moreover, \cite{Reddy2014Icar..237..116R} propose that the Baptistina asteroid family is the source of LL chondrites that show shock blackened impact melt material. 

The NEA population also contains a good amount of X-complex asteroids \citep{Binzel2015aste.book..243B}, which origin remains unclear, despite their link with meteorites and asteroids being visited by space missions (see section~\ref{S:Xcomplex}). 

This work focuses on the search for X-types families in the inner Main Belt, which could represent sources of X-type NEAs, in particular of those with intermediate albedo. \cite{Dykhuis2015Icar..252..199D} has already noted the presence of a small X-type family, that they called Hertha-2, that could produce some NEAs. However, this latter family is arguably too small to account for the flux of the observed X-type NEAs. 
In section~\ref{S:Xcomplex} we describe the main physical properties of X-complex asteroids; in section~\ref{S:Xsource} we analyse the source regions of X-type NEAs and we show that the inner Main Belt has significantly higher probability compared to other areas of the Solar System to deliver X-type asteroids to the near-Earth space; in sections~\ref{S:methods} and \ref{S:results} we describe our search and identification of families amongst X-type asteroids of the inner Main Belt, while in section~\ref{S:discussion} we present the implications of our findings.

\section{Characteristics of the spectroscopic X-complex \label{S:Xcomplex}}

\begin{figure}
\centering
\includegraphics[width=\columnwidth]{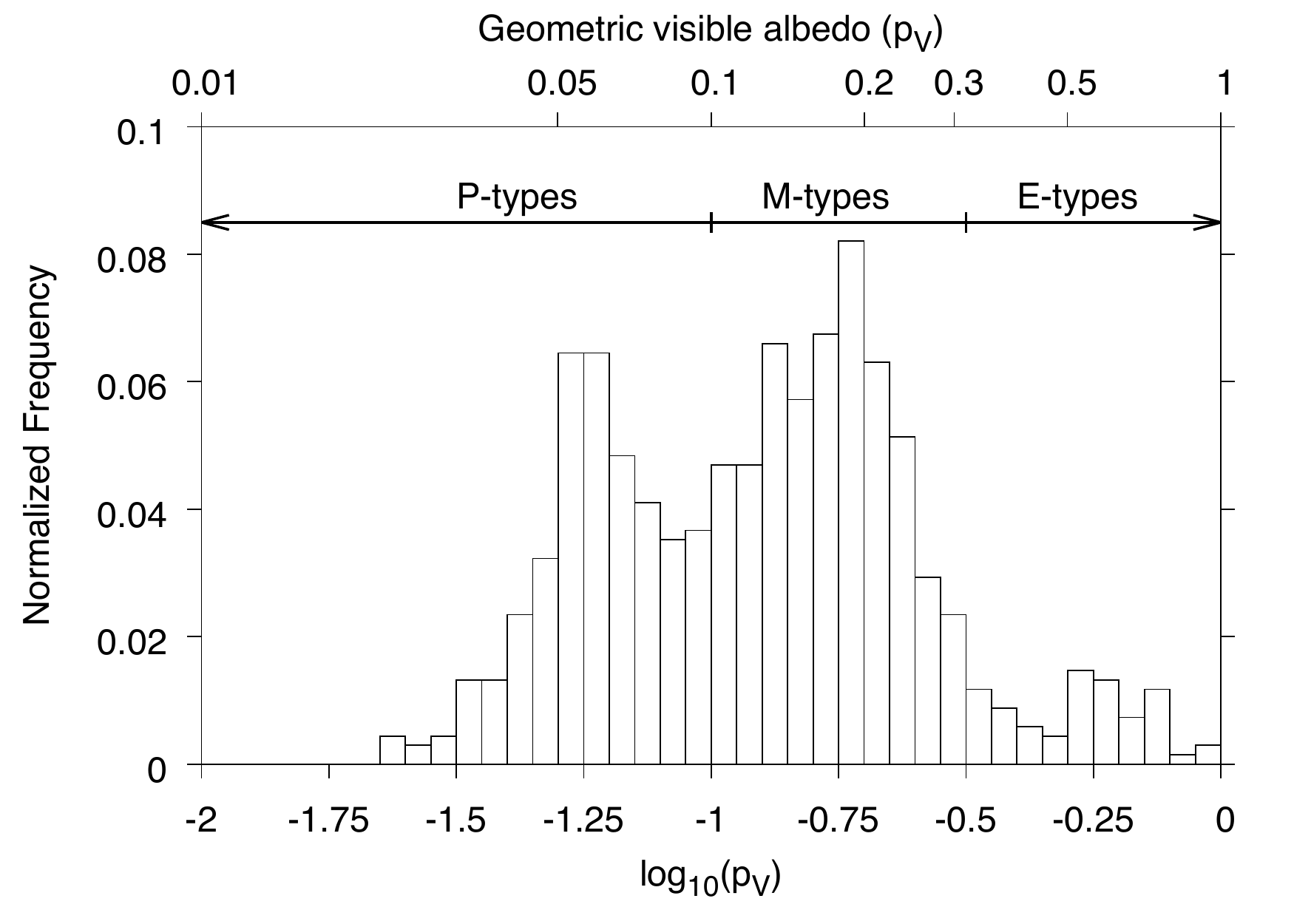}
\caption{\label{globalXpVdist} \ca{The multimodal distribution of the geometric visible albedo of all asteroids belonging to the X-complex (i.e. with spectral types found in literature being X, Xc, Xe, Xk, M, E, and P). The three peaks of the distribution correspond to the P-, M-, E-types of the \cite{Tholen1989aste.conf..298T} taxonomy, and their boundaries are defined respectively $p_V\leq 0.1$, $0.1 \leq p_V \leq 0.3$, and $p_V>0.3$, and are marked by the arrows at the top of the figure. As one can see these albedo boundaries provide reasonable separation between the classes. In this work we focus on X-complex asteroids with $0.1 \leq p_V \leq 0.3$ i.e. the M-types of \cite{Tholen1989aste.conf..298T}, which in Bus-DeMeo is mostly corresponding to the X$_c$-types.}}
\end{figure}

The spectroscopic X-complex is characterised by moderately sloped spectra with no or weak features and it is compositionally degenerate, as it contains objects with high, medium and low albedos \citep{Fornasier2011Icar..214..131F,DeMeo2015aste.book...13D}. For instance, the X-complex in the Tholen taxonomy \citep{Tholen1989aste.conf..298T} is primarily separated into the E-, M- and P-types which have different albedo ranges (see Fig.~\ref{globalXpVdist}). According to the more recent Bus-DeMeo taxonomy \citep{DeMeo2009Icar..202..160D}, the X-complex contains the Xe, Xc and Xk classes with very different inferred mineralogies. Spectroscopically, these classes are very close to some C-complex classes, such as the Cg, Ch and Cgh, and thus more detailed analysis of their specific features is needed \citep{DeMeo2009Icar..202..160D}. The low-albedo X-complex asteroids could be compositionally similar to those of the C-complex \citep{DeMeo2009Icar..202..160D,DeMeo2015aste.book...13D}. An indication of the composition similarities and origin between the low-albedo asteroids belonging to the X- and C-complex also comes from the fact that C- and X-complex asteroids have been found within the same families \citep{Morate2016A&A...586A.129M,Fornasier2016Icar..269....1F}. In the near-infrared  survey of \cite{Popescu2018A&A...617A..12P} asteroids belonging to the X-complex are denoted by the class Xt. 

The high albedo E-types ($p_V>0.3$) are represented by the Xe-type asteroids and have a flat spectrum with a weak absorption band at 0.9~$\si{\um}$ and a deeper one at 0.5~$\si{\um}$ \citep{DeMeo2009Icar..202..160D}. They have been compositionally linked to the enstatite achondrite (aubrites) meteorites \citep{gaffey1992,fornasier2008} and their main reservoir at small heliocentric distances is found in the Hungaria region at \ca{high inclination \citep{gaffey1992,Cuk2014Icar..239..154C};} in particular they are members of the Hungaria family, which is superimposed on an S-complex dominated background \citep{lucas2017}. 

The traditional \cite{Tholen1989aste.conf..298T} M-type group, with moderate albedo range ($0.1<p_V<0.3$), has been shown to contain objects with several compositions, including metallic objects of iron/nickel composition, thought to be the parent bodies of the iron meteorites and supposed to represent the cores of differentiated objects. In the Bus-DeMeo taxonomy moderate albedo asteroids belong to the Xk and Xc classes. In particular, it has been shown that asteroids classified as Xk are compositionally linked to the mesosiderite meteorites; in particular with asteroids (201) Penelope, (250) Bettina and (337) Devosa \citep{vernazza2009}. Xc-type asteroids having a reflectance spectrum with the shallowest slope compared to the rest X-complex asteroids in the range 0.8–2.5~$\si{\um}$ constitue the only asteroid class of the X-complex that is characterised by the absence of the 0.9~$\si{\um}$ feature, the latter linked to the presence of orthopyroxene \citep{hardersen2005}. Asteroids of the Xc-class have been linked to the enstatite chondrite (EC) meteorites. Two characteristic Xc-types are (21) Lutetia and (97) Klotho \citep{vernazza2009}.

On the other hand, P-types, the dark asteroids of the X-types, are not well represented in the newer taxonomy. They are located mainly in the outer belt and are similar to C-complex asteroids, being also linked with CM meteorites \citep{Fornasier2011Icar..214..131F}.

From the current available data there are two Xk, (56) Melete and (160) Una, and one Xc asteroid, (739) Mandeville, that have very low albedo values, showing that there is no strict matching between the medium albedo M-types and the Xc/Xk -ypes.

\section{Source regions of X-type NEAs \label{S:Xsource}}

For our study we extract asteroid information from the Minor Planet Physical Properties Catalogue (\url{mp3c.oca.eu}), developed and hosted at Observatoire de la C\^ote d'Azur. This database contains orbits and physical properties of both MBAs and NEAs. Especially, for NEAs it also includes albedo and spectral classes obtained from the Data Base of Physical and Dynamical Properties of Near Earth Asteroid of the E.A.R.N. (European Asteroid Research Node) hosted by the DLR Berlin (\url{http://earn.dlr.de}).

We select NEAs of the X, Xk, Xc, Xe, E, M, P-types according to Tholen \citep{Tholen1989aste.conf..298T}, Bus \citep{Bus2002Icar..158..106B} and Bus-DeMeo \citep{DeMeo2009Icar..202..160D} taxonomies. As a second filter we require their geometric visible albedo ($p_V$) to be in the range $0.1<p_V<0.3$. 
This albedo cuts exclude the bright Xe-types, mostly linked to the Hungaria region, and the dark X-type population probably linked to carbonaceous asteroids. \ca{We note that P-types had, by definition, $p_V <$~0.1 in the taxonomy of \cite{Tholen1989aste.conf..298T}. However, some of the $p_V$-values have been revised after the original Tholen's work, making it possible that some P-types  could have revised $p_V >$~0.1. However, we find no such cases.}

\begin{figure}
\centering
\includegraphics[width=\columnwidth]{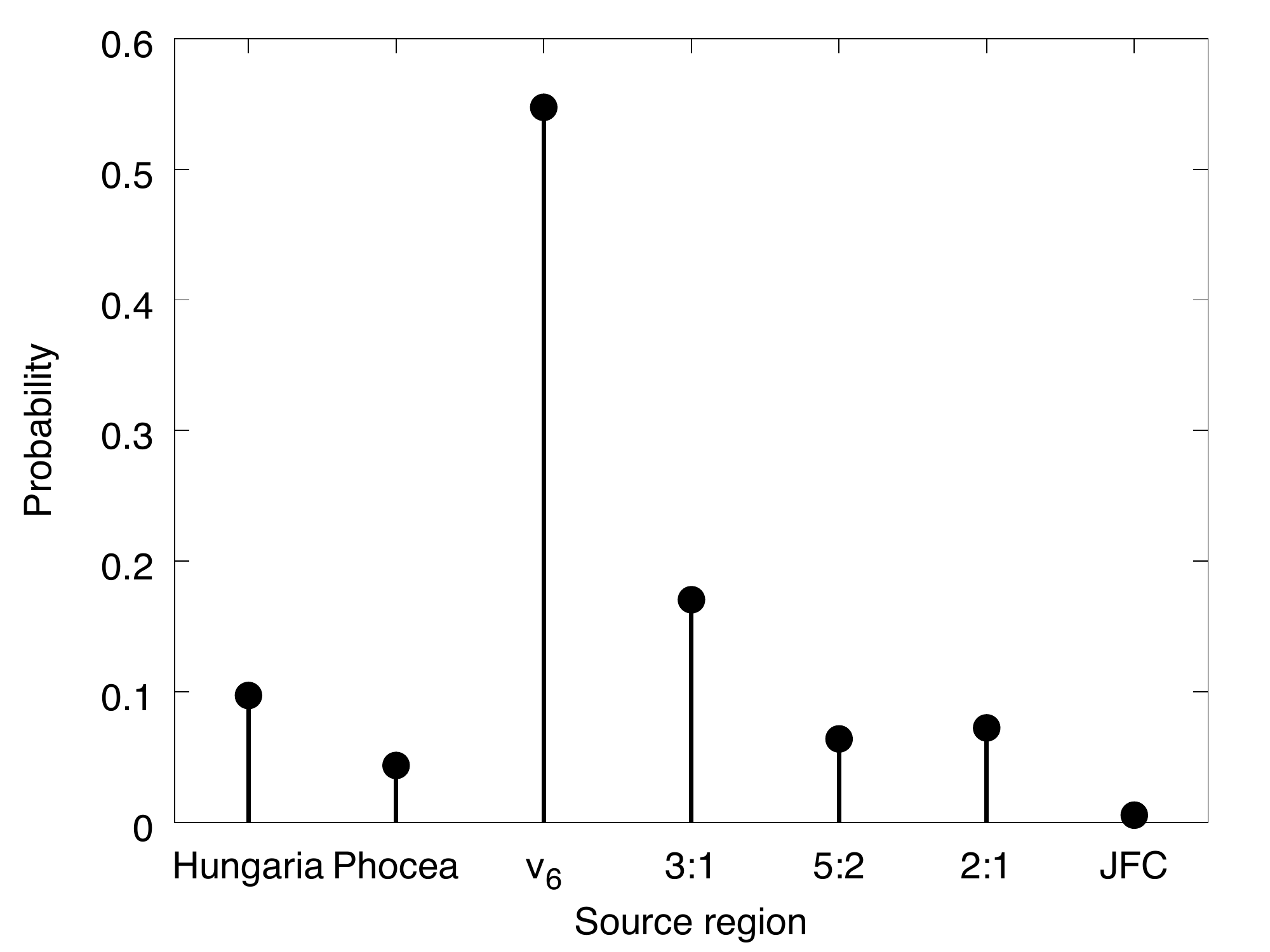}
\caption{\label{F:XtypeNEAsource} The source regions of the 15 selected NEA which fulfil the criteria in spectral class and albedo range. The region with the highest probability on average is the $\nu_6$, indicating that the source should be the inner Main Belt.}
\end{figure}

For each NEA that passes our selection criteria -- 15 asteroids in total 
-- we extract its source region probabilities, which we take from the work of \citep{Granvik2016Natur.530..303G,Granvik2017A&A...598A..52G}. In this model there are seven source regions for the NEAs; the Hungaria (HU) and Phocaea (PHO), the J3:1, J5:2 and J2:1 MMR, the Jupiter-family comets (JFC) and the $\nu_6$ complex, the later including the $\nu_6$ secular resonance, the J4:1 and J7:2 MMR.
We take the mean of the source probabilities for the considered NEAs and display the results in Fig.~\ref{F:XtypeNEAsource}, which shows that the $\nu_6$ source dominates, with the second most effective source region being the J3:1. \ca{The latter could also contribute with asteroids drifting inward from the central Main Belt ($2.5<a<2.82$~au)}. The J4:1 MMR and the  $\nu_6$ secular resonance determine the inner border of the Main Belt, while the J7:2 that overlaps with the M5:9 MMR, at heliocentric distance of 2.256~au, also delivers to near-Earth space asteroids from the inner portion of the Main Belt \citep[see also][]{Bottke2007Natur.449...48B}.

The average size (diameter) of the selected NEAs, calculated around 1.5~km, implies that these bodies can hardly be planetesimals that formed 4.567~Gyr ago. This is because their size dependent collisional lifetime is $<1$~Gyr \citep{Bottke2005Icar..175..111B}. In addition, there is evidence that the planetesimals, i.e. the original asteroids, formed much bigger, possibly with sizes around 100~km \citep{Morbidelli2009Icar..204..558M}, with shallow size distribution \citep{tsirvoulis2018} and certainly bigger than few tens of km in diameter \citep[35~km, see][]{Delbo2017Sci...357.1026M}. The aforementioned argument implies that these NEAs originate from a more recent fragmentation of a larger parent that formed a family. Slowly the family members drifted by the Yarkovsky effect into one of the source regions, which removed them from the Main Belt and delivered them to the near-Earth space. 

The question that arises is which are the potential families that feed the NEA population with X-complex objects with moderate albedo. The most well known X-complex family is Hungaria in the Hungaria region. However, as described earlier, this has several Xe-type asteroids with high albedo values, $p_V>0.3$ that have been discarded by our filtering criterion. In the inner Main Belt there are five families that are characterised in the literature as X- or CX-type: Clarissa, Baptistina, Erigone, Chimaera and Svea \citep{Nesvorny2015aste.book..297N}. A closer inspection of the family members of Clarissa (introduced as an X-type) shows that they have very low albedos (average family albedo is $p_V=0.05$)  consistent with a P-type classification in the \cite{Tholen1989aste.conf..298T} taxonomy. The asteroid (302) Clarissa itself is classified as an F-type, and in visible wavelengths is spectroscopically similar to (142) Polana (another F-type), the potential parent body of the Polana family \citep{Walsh2013Icar..225..283W}. Therefore, it is more appropriate to consider Clarissa as a family with a carbonaceous composition. 
The Erigone family has an average albedo of $p_V=0.06$, while the few objects that have been classified as X or CX have very low albedos ($p_V<0.08$) which indicate that also this family has very likely a carbonaceous composition. 
Likewise, Svea has also a carbonaceous composition, because the average and standard deviation of the albedo distribution of its members are $p_V$~=~0.06 and 0.02, respectively. 
The only member of the Svea family classified as an X-type is (13977) Frisch that has $p_V$=0.16 $\pm$ 0.02, which is 5$\sigma$ from the mean. This could be an indication that this asteroid is an interloper and in reality does not belong to Svea family. 
Chimaera, with average $p_V=0.07$ and a standard deviation of 0.05, has 108 asteroid members of which eight have been indicated either as CX or as C and X (in different taxonomies). This indicates that also this family cannot  produce moderate albedo X-type NEAs.

The Baptistina asteroid family is young \citep[$t<300$~Myr,][]{broz2013}, contains almost 2,500 members \citep{Nesvorny2015aste.book..297N} of moderate albedos (mean $p_V=0.16$), and is capable of delivering objects in the near-Earth space \citep{Bottke2007Natur.449...48B}. 
According to spectroscopic and spectrophotometric observations, 417 family members have been classified of which 154 as S-types, 172 as C, CX (or simultaneously C and X), while only 18 X-types. The asteroid (298) Baptistina itself is spectroscopically classified as an Xc-type object. A recent study has classified (298) Baptistina as S-type and suggests that it is the parent body of the Chelyabinsk bolide \citep{Reddy2014Icar..237..116R}, which hit Earth's atmosphere on February 13, 2013. Chelyabinsk meteorites, which resulted from the bolide, match the composition of the LL ordinary chondrite meteorites. In general, LL chondrites are linked to S-type asteroids, and the most prominent family spectroscopically similar to LL chondrites is Flora. However, analysis of 11 members of the Baptistina family show spectra alike those of LL5 chondrites, but with lower albedos compared to typical S-types. This lower albedo can be explained by the presence of blackening impact melts \citep{Reddy2014Icar..237..116R}. Baptistina and Flora families overlap in the orbital element space and it is an open question whether the Baptistina family was formed by the breakup of a once-upon-a-time Flora member \citep{Nesvorny2015aste.book..297N}.

From the above considerations, there is no solid evidence for the presence of an inner Main Belt X-type family with moderate albedos being also the source of  Xc/Xk NEAs. Since moderate albedo X-complex asteroids have been observed in near-Earth space and originate from the-inner Main Belt it is possible that one or more diffused families of X-types with moderate albedos have been so far escaped identification by classical family searching methods such as the HCM. In the following, we use a new method \citep{Bolin2017Icar..282..290B}, already successfully tested \citep{Delbo2017Sci...357.1026M}, to search for the missing X-type families in the inner Main Belt.

\section{Materials and methods}
\label{S:methods}

\subsection{Selection of asteroids \label{S:inputPopulation}}
We restrict our search for the intermediate albedo component of the X-complex (see Fig.~\ref{globalXpVdist}) and thus we select all asteroids of the inner Main Belt (with 2.1 $< a <$ 2.5 au) with $0.1 \leq p_V \leq 0.3$ belonging to the X-complex, having spectral taxonomic classes of X, Xc, Xe, Xk, Xt, M, E, as described in section~\ref{S:Xcomplex}. \ca{We also check for  \cite{Tholen1989aste.conf..298T} P-types in the IMB, which could have a revised $p_V>0.1$, finding no such cases. This selection resulted in a pool of 386 asteroids.}


We use the database of \cite{Delbo2017Sci...357.1026M}, which can now be accessed from the \url{mp3.oca.eu} (see section~\ref{S:Xsource}) to perform these extractions. The database is used to extract proper orbital elements and diameter values for the selected asteroids on which we apply the family searching technique. In this database, all known asteroids as of 2016 November 4 from the Minor Planet Center (\url{https://www.minorplanetcenter.net}, downloaded Nov 2016) are crossmatched with (i) the synthetic and analytic (for those asteroids without synthetic) proper elements \cite{Milani2014Icar..239...46M} of the AstDys-2 database (\url{http://hamilton.dm.unipi.it/astdys/}); (ii) radiometric diameters and albedos taken from \cite{Nugent2016AJ....152...63N,Nugent2015ApJ...814..117N,Masiero2012ApJ...759L...8M,Tedesco2002AJ....123.1056T,Ryan2010AJ....140..933R,Usui2011PASJ...63.1117U,Masiero2011ApJ...741...68M,Masiero2014ApJ...791..121M} 
combined in reverse order of preference (i.e. results from a following-cited reference overwrite those of a preceding-cited one, in order to have a unique diameter and albedo value for each asteroid);
(iii) spectral classification from several sources
\cite{NeeseTaxPDS,Carvano2010A&A...510A..43C,Demeo2013Icar..226..723D,deLeon2016Icar..266...57D}; 
and (iv) rotational periods from the Asteroid Lightcurve database \cite{lcdb}.

\subsection{Family search method\label{S:VshapeMethod}}
Since our working hypothesis is that a putative missing X-complex family is dispersed and therefore very likely old, we use the method of \cite{Bolin2017Icar..282..290B} to search for Yarkovsky V-shapes of families in the inner Main Belt. This technique searches for V-shapes of unknown age and vertex in an asteroid population, in the 2-dimensional space of parameters representing the centre of the family, $a_c$, and the slope, $K$, of the sides of the V-shape. To do so, the method draws in the $(a,1/D)$ space a nominal-V described by the equation $1/D = K | a - a_c|$, an inner-V with equation $1/D = (K- \Delta K) | a - a_c|$, and a outer-V with equation $1/D = (K + \Delta K) | a - a_c|$. Next, it counts how many asteroids fall in between of the borders of the inner- and the nominal-V and between the borders of the nominal- and the outer-V, indicating $N_\text{in}$ and $N_\text{out}$ these numbers, respectively. It then plots the value of $N_\text{in}^2/N_\text{out}$ as a function of $a_c$ and $K$. Local maxima of $N_\text{in}^2/N_\text{out}$ indicate V-shaped over-densities of asteroids which may be associated to a Yarkovsky-evolved asteroid family

In order to determine the uncertainties on the values of $K$ and $a_c$, we improve the technique used by \cite{Delbo2017Sci...357.1026M}, who adopted the method of \cite{Spoto2015Icar..257..275S}, by using here a more rigorous approach. \ca{First of all, since our selection of asteroids depends on the values of their geometric visible albedo to be within the aforementioned limits (0.1~$\leq p_V \leq$~0.3), we perform 10$^4$ Monte Carlo simulations where the nominal values of the albedos and diameters of asteroids are varied within their uncertainties, which are assumed to be 1$\sigma$ values drawn from Gaussian distributions. At each iteration, a new selection is performed based on the albedo range and} a new V-shape search is performed around the nominal centre and slope of the V-shape and the values of $a_c$ and $K$ of each iteration that maximise $N_\text{in}^2/N_\text{out}$ are recorded. The RMS of the distributions of the recorded $a_c$ and $K$ are taken as the uncertainties on those parameters.

\subsection{Statistical tests}
\label{S:StatTest}
Following the method of \cite{Delbo2017Sci...357.1026M} we assess whether a V-shape distribution of asteroids could be due to statistical sampling instead of indicating the existence of a real V-shaped over density, diagnostic of an asteroid family.
To do so, we perform a statistical test where we assume the null-hypothesis that the distribution of the values of the proper semimajor axis is size independent. This null-hypothesis would mean that the asteroids are neither genetically related nor dispersed by the Yarkovsky effect from a common centre. Then, for each asteroid we preserve its diameter value and assign a new proper semimajor axis. This value of $a$ is randomly extracted each time from the proper semimajor axis distribution of all X-complex asteroids of the inner Main Belt with 0.1 $< p_V <$ 0.3 regardless of their size.
We typically perform 10$^6$ iterations and we measure in which fraction of the trials we generate a V-shape as the one observed, i.e. what is the fraction of the simulations with a number of asteroids falling outsize our nominal V-shape borders that is smaller or equal to the observed one.
If we find, for instance, that in 68.27\% or 99.73\% of the simulations the number of asteroids outsize the V-shape borders is smaller or equal to the observed one, this implies that the existence of the family is robust at 1 and 3$\sigma$ respectively.

\subsection{Identification of a family core}
\label{S:Core}
Once a V-shape is located (as e.g. in Fig.~\ref{F:vshapeScan}), we also attempt to determine whether the distributions of the orbital eccentricity and inclination of the asteroid population near and inside the borders of the V-shape have peaks, possibly indicating orbital clustering of the family members. 
To do so, we plot the distribution of the number of asteroids in bins of eccentricity and inclination normalised by the amount of orbital phase space available in each bin. 
For the case of the eccentricity, the available $a$-space between the Mars-crossing boundary and the J3:1 MMR decreases with increasing $e >$~0.2. This is because asteroids with $e >$~0.2 and 2.1$<a<$ 2.5 au can have perihelion distances $q=a(1-e)$ below the aphelion distances $Q_{Mars}=$1.666 au of Mars. For the lower ($e_{l,j}$) and upper limit ($e_{h,j}$) of each $j^{th}$-bin of $e$, we compute their corresponding $a$-values ($a'_{l,j}$ and $a'_{h,j}$) using Eq.~\ref{E:eccMars} that represent the Mars-crossing curve:
\begin{equation}
\begin{aligned}
a' =  2.1                            ~~~& \text{for}~e \leq 1-Q_{Mars}/2.1,\\
a' =  1-Q_{Mars}/(1-e)      ~~~& \text{for}~e > 1-Q_{Mars}/2.1.
\label{E:eccMars}
\end{aligned}
\end{equation}
We then compute the area of the trapezium-shaped $a$ space between the Mars-crossing curve and the J3:1 MMR of the inclination bin: 
\begin{equation}
area'_j = (a'_{l,j} - 2.5 + a'_{h,j} - 2.5) \times (e_{h,j} - e_{l,j}) /2. 
\end{equation}
\ca{Figure~\ref{F:agapiAEIvel} shows that the distribution of X-type asteroids with 0.1~$\leq p_V \leq$~0.3 of the IMB is probably unaffected by the Mars-crosser border, but we prefer to include it for sake of generality of the method.}
For the case of the inclination, the available $a$-space between the $\nu_6$ and the J3:1 MMR decreases with increasing inclination, as the $a$-value of centre of the $\nu_6$ varies from $\sim$2.1 au at $\sin i$ = 0 to 2.5 au for $\sin i$ = 0.31. We approximate the shape of the $\nu_6$ with a second order polynomial of the form:
\begin{equation}
a = A \sin^2 i + B \sin i + C,
\label{E:nu6}
\end{equation}
where $A,B,C$ have the values of 4.99332, -0.287341, 2.10798  au, respectively. 
\ca{We note that the location of a secular resonance depends also on eccentricity. However, for the case analysed in this work, all asteroids have $\sin i \lesssim$ 0.3 (i.e. $i \lesssim$ 17$^\circ$). The $\nu_6$ secular resonance at these inclinations or smaller is essentially eccentricity independent \citep[it is almost a straight line as shown by Fig.~6 of][]{Morbidelli1991CeMDA..51..131M}. On the other hand, for other applications at much higher eccentricities, 
different approximations of the path of the  $\nu_6$ could be needed as function of the eccentricity.}

For the lower ($\sin i_{l,j}$) and upper limit ($\sin i_{h,j}$) of each $j^{th}$-bin of $\sin i$, we compute their corresponding $a$-values ($a_{l,j}$ and $a_{h,j}$) using Eq.~\ref{E:nu6}; we then compute the area of the trapezium-shaped $a$ space between the $\nu_6$ and the J3:1 MMR of the inclination bin: 
\begin{equation}
area_j = (a_{l,j} - 2.5 + a_{h,j} - 2.5) \times (\sin i_{h,j} - \sin i_{l,j}) /2. 
\end{equation}
%
For each bin in $e$ (and $\sin i$), we count all bodies in a narrow sliver inner of the nominal V-shape, i.e. with $(a,1/D)$ verifying the condition
$K | a - a_c| \leq 1/D \leq K | a - a_c| + W$ (where $W$ is small fraction, e.g. 10--20\%, of the value of $K$)
and whose values of $e$ (and $\sin i$) are included in in each bin. We then multiply the number of asteroid in each bin by the value of $e$ (and $\sin i$) of the center of the bin and divide by the bin area. The number obtained for each bin is normalised by the total number of asteroid multiplied the total area. The multiplication by the value of $e$ and $\sin i$ for the eccentricity and the inclination histograms, respectively, is due to the fact that the orbital phase space available to asteroids is proportional to $e$ and $\sin i$.

If the histograms of the normalised $e$ and $\sin i$ distributions have well defined peaks (see Fig.~\ref{F:incEccDistrib})  
 -- which was not the case for the primordial asteroid family discovered by \cite{Delbo2017Sci...357.1026M} -- we denote by $e_c$ and $i_c$ the coordinates of the peak in eccentricity and inclination, respectively. 
 
 Next, we use the Hierarchical Clustering Method \citep[HCM][]{Zappala1990AJ....100.2030Z} to determine the population that clusters around the family centre. To do so, we create a fictitious asteroid with orbital elements ($a_c$, $e_c$, $i_c$) that we use as central body for the HCM algorithm. We use the HCM version of  \cite{Nesvorny2015aste.book..297N}. 
We follow this procedure to avoid using real asteroids near the vertex of the family V-shape. This is because for old and primordial families, it is possible that the parent asteroid was lost due to the dynamical evolution of the Main Belt \citep{Delbo2017Sci...357.1026M,Minton2010Icar..207..744M}. 

The HCM algorithm relies on the "standard metric" of 
\citep{Zappala1990AJ....100.2030Z,Zappala1995Icar..116..291Z}
to calculate a velocity difference between two asteroid orbits. It then connects bodies falling within a cut-off velocity ($V_c$),  creating clusters of asteroids as a function of $V_c$. The number of asteroid in the cluster $N_c$ is a monotonic growing function of $V_c$, but as this latter increases the number of family members can grow with steps. For small cut-off velocities no asteroids are linked; when $V_c$ reaches a value appropriate to cluster members of the family, the number of members jumps up, followed by a steady increase of $N_c$ with further increasing value of $V_c$, until all family members are linked to each other; if one keeps increasing $V_c$, the value of $N_c$ jumps up again followed by another slow increase. This second jump indicates the limit of the family, because after the jump, the HCM accretes objects outside the family or asteroids belonging to adjacent families. We use the value $V_c$ at the second jump of $N_c$ to define the family. More precisely, we run the HCM for values of $V_c$ between 10 and 600~m~s$^{-1}$ with a step of 1~m~s$^{-1}$ and we take as nominal $V_c$ of the family the minimum value of this parameter that allows to link the largest number of asteroids before the second jump in the value of $N_c$.

\subsection{Family age determination}\label{S:age}
In order to provide an estimate for the age of the family $T$, and its uncertainty, we use two methods.
Firstly, we follow the procedure of \cite{Delbo2017Sci...357.1026M}, which is based on the methods of \cite{Spoto2015Icar..257..275S} that gives an approximate age for the family (as this technique does not take into account the effect of YORP cycles on the Yarkovsky drift of asteroids). 
The age ($T$) and its uncertainty are derived from the inverse slope of the V-shape given by Eq.~\ref{E:InvSlopeAge}:
\begin{equation}
1/K = (da/dt)_\text{1km}~T, 
\label{E:InvSlopeAge}
\end{equation}
where (d$a$/dt)$_\text{1km}$ is the rate of change of the orbital proper semi-major axis with time (t) for an asteroid of 1~km in size due to the Yarkovsky effect \citep{Bottke2006AREPS..34..157B}.
We perform 10$^6$ Monte Carlo simulations, where at each iteration, random numbers are obtained from the probability distributions of $K$ and (d$a$/dt)$_\text{1km}$ in order to be used in Eq.~\ref{E:InvSlopeAge}.
The (d$a$/dt)$_\text{1km}$ value, obtained by applying the formulas of \cite{Bottke2006AREPS..34..157B}, is calculated at each Monte Carlo iteration, and depends on asteroid properties such as the heliocentric distance, infrared emissivity, bolometric Bond albedo ($A$), thermal inertia, bulk density, rotation period, and  obliquity ($\gamma$), which is the angle between the asteroid's spin vector and its orbital plane. The border of the V-shape is determined by those asteroids drifting with maximum $|da/dt| \propto \cos \gamma$, so that we assume $\gamma = \pm 90^\circ$.  For the other parameters, we extract random numbers from their probability distributions. The calculation of the probability distributions for each of the relevant parameter is described in section~\ref{S:results}. A change in the Sun's luminosity as function of time is taken into account using Eq.~4 of \cite{Carruba2015MNRAS.451..244C}.

\section{Results}
\label{S:results}

\begin{figure}[h!]
\includegraphics[width=\columnwidth]{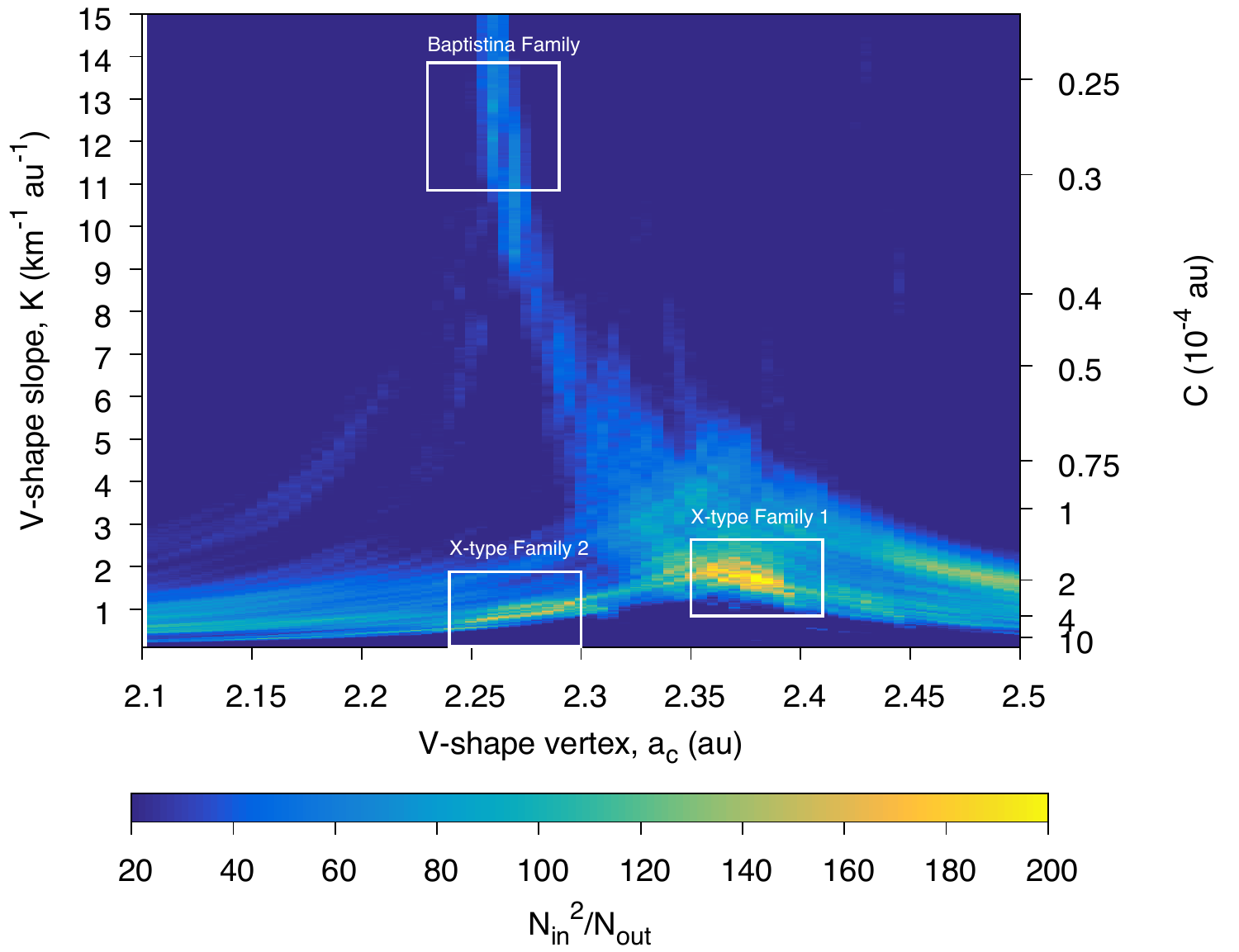}
\caption{\label{F:vshapeScan} Output of the V-shape searching method. 
The value of $N_\text{in}^{2}/N_\text{out}$ is plotted as a function of the slope ($K$) and semimajor axis of the vertex of a V-shape ($a_c$). The value of the parameter $C$, \ca{which measures the width of a V-shape in the $(a,H)$ space, where $H$ is the absolute magnitude} \citep{Vokrouhlicky2006Icar..182..118V}, is calculated from $C = 1/K \sqrt{p_V} /1329$ using a geometric visible albedo $p_V$~=~0.2.}
\end{figure}

We use the V-shape searching technique on all moderate albedo X-types of the inner Main Belt selected as detailed in section~\ref{S:inputPopulation}. The result of the V-shape search is shown in Fig.~\ref{F:vshapeScan}, where we identify a prominent peak of $N_a^2/N_b$ at $a_c$=2.38 au and $K$=1.72 km$^{-1}$ au$^{-1}$, which we identify with the label "1" in the figure. Other two peaks of $N_a^2/N_b$ at $a_c\sim$2.28 au and $K\sim$1 km$^{-1}$ au$^{-1}$ (peak "2" of Fig.~\ref{F:vshapeScan}) and at $a_c$=2.26 au and $K$=11.9 km$^{-1}$ au$^{-1}$ are also visible. The latter peak corresponds to the V-shape of the Baptistina asteroid family \citep{Nesvorny2015aste.book..297N,Bottke2007Natur.449...48B,Masiero2012ApJ...759...14M}, while the first and second peak represent previously unknown V-shapes 
(Fig.~\ref{F:vshape} gives a graphical representation of the V-shapes).

We begin focusing on the V-shape centred at  $a_c$=2.38. In order to determine the uncertainties on the values of its $a_c$ and $K$ parameters, we perform a Monte Carlo simulation as described in section \ref{S:VshapeMethod} (the ranges of parameter values explored is $1 \leq K \leq 4$ au$^{-1}$~km$^{-1}$ and $2.34 \leq a_c \leq 2.40$~au). We find that the \ca{uncertainties on the determination of $a_c$ and $K$ are 0.006 au and 0.1~au$^{-1}$~km$^{-1}$}, respectively. \ca{We note that the average catalogue albedo uncertainty for our pool of asteroids is about 30\% relative value, which is similar to the albedo uncertainty estimated by \cite{Pravec2012Icar..221..365P} for the WISE catalogue.}

\begin{figure*}[ht]
\includegraphics[width=\columnwidth]{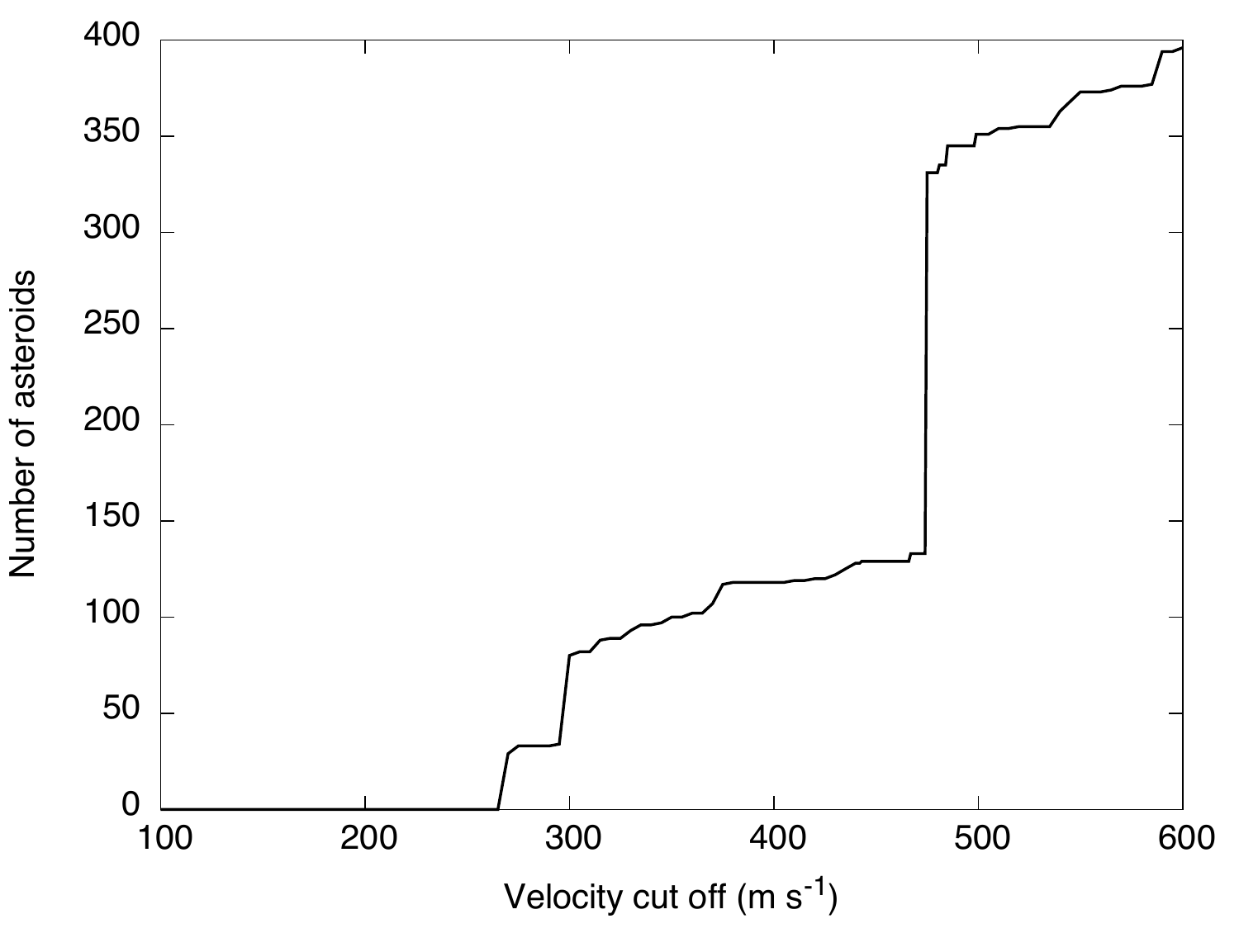}
\includegraphics[width=\columnwidth]{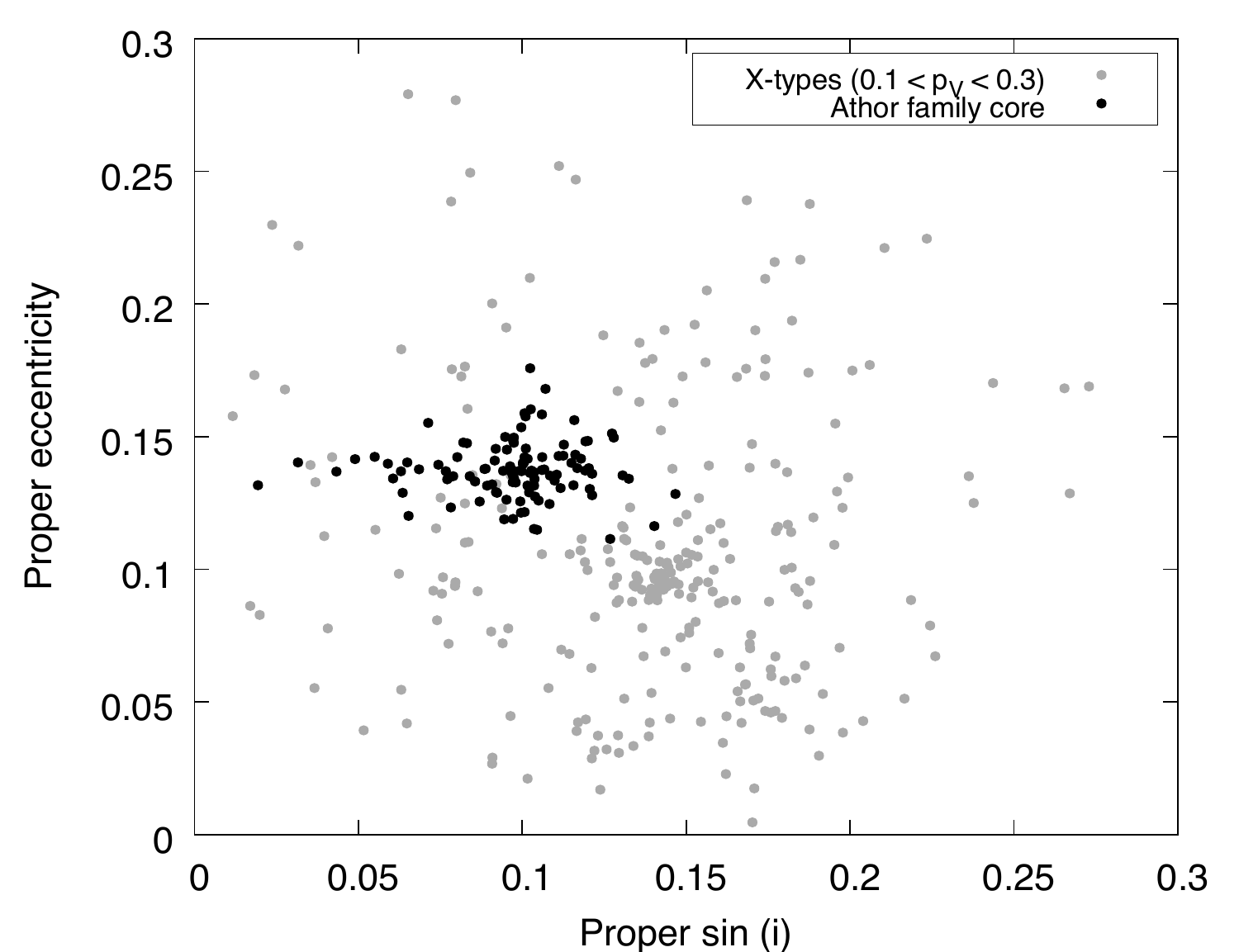}\\
\includegraphics[width=\columnwidth]{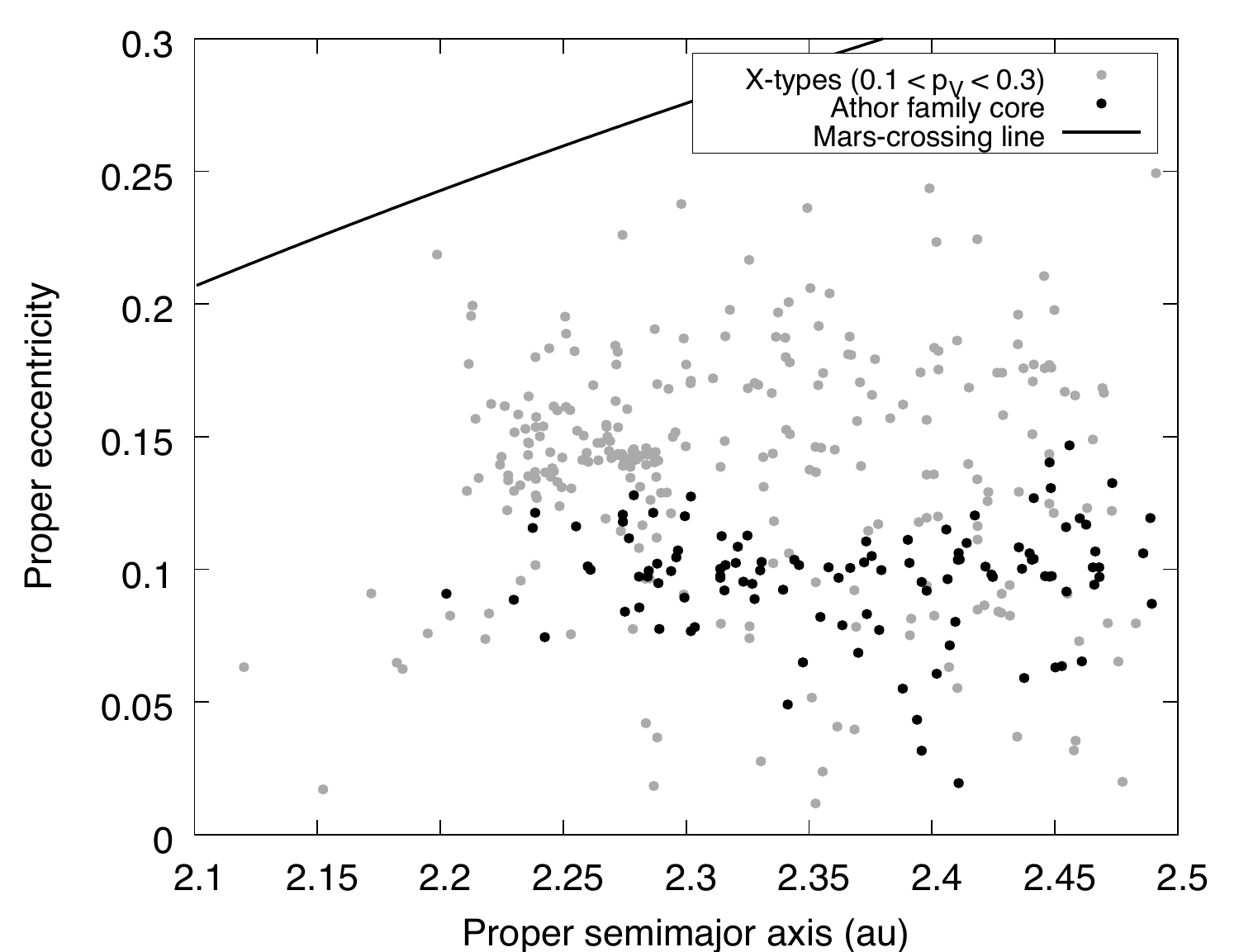}
\includegraphics[width=\columnwidth]{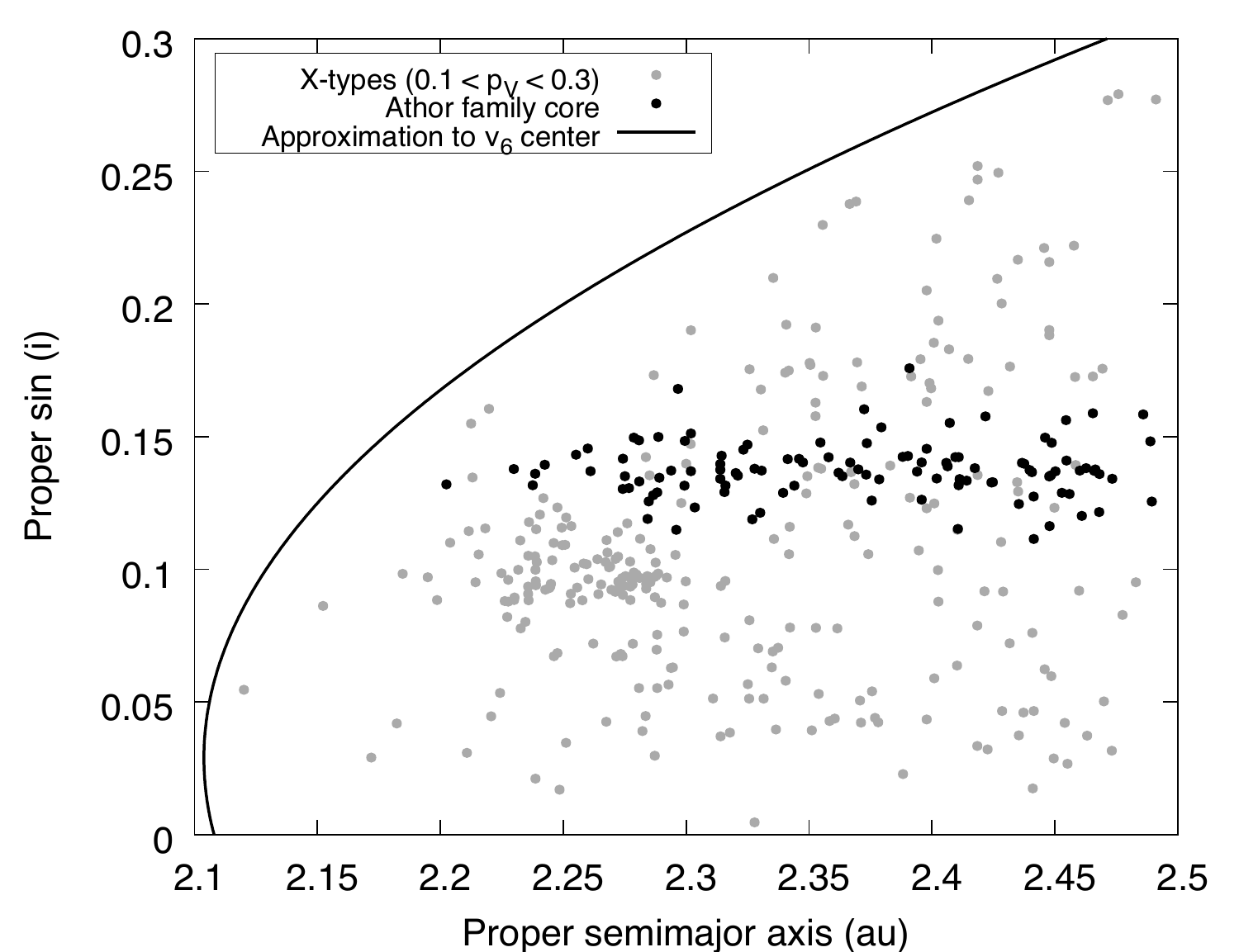}\\
\caption{\label{F:agapiAEIvel}
{\it Top-left panel:} The number of family core members found as a function of velocity cutoff, using the HCM. The second jump in the number of linked asteroids, at 460~m~s$^{-1}$, is our adopted value of velocity cutoff.
{\it Other panels:}  Distribution of the X-complex asteroid with 0.1~$<p_V<$~0.3 of the inner-Main Belt in the three proper elements $a$, $e$, $\sin i$. The cluster (of grey points) centred at $(a, e,~\sin i) \sim (2.28~au, 0.144, 0.095)$ is the core of the Baptistina asteroid family.}
\end{figure*}

Next, we study the inclination and eccentricity distributions of those asteroids inside and near the border of the V-shape centred at ($a_c$, $K$)~=~(2.38 au, 1.72 km$^{-1}$ au$^{-1}$), as detailed in section \ref{S:VshapeMethod}. These distributions, presented in Fig.~\ref{F:incEccDistrib}, show maxima at $e_c$~=~0.12 and $\sin(i_c)$~=~0.14, indicating a potential clustering of objects in $(e, i)$ space. Together with the vertex of the V-shape, this information points to a putative centre of the family at $(a_c, e_c, \sin i_c)$~=~(2.38 au, 0.12, 0.14). 
We assume a fictitious asteroid with the above-mentioned proper orbital elements for the central body for the HCM algorithm. The latter is used to link asteroids as a function of $V_c$. Again, we use moderate albedo X-type asteroids of the inner Main Belt selected as described in section~\ref{S:inputPopulation}, as input population for the HCM. The number of asteroids linked by the HCM is displayed in Fig.~\ref{F:agapiAEIvel}, which shows that $V_c \sim$ 460 ~~s$^{-1}$ identifies the second jump in the number of bodies linked to the family as function of  $V_c$. We thus take this value as the nominal $V_c$ for this cluster. The lowest-numbered and largest asteroid of this cluster is the M-type (161) Athor, which is also the asteroid closest to the vertex of the V-shape. 
%

Figure~\ref{F:agapiAEIvel} shows that that X-type asteroids of the inner Main Belt have bimodal distributions in $e$ and $\sin i$ with the cluster identified by the HCM and parented by (161) Athor located at high inclination ($\sin i >$~0.1), and low eccentricity ($e\sim$~0.1). Another very diffused group of asteroids appears at low inclination and high eccentricity ($\sin i <$~0.1, $e >$~0.12). Once members of the Baptistina family are removed, most of X-type asteroids with $\sin i <$~0.1 and $e >$~0.12 are included inside the V-shape defined by ($a_c$, $K$)~=~(2.38 au, 1.72 km$^{-1}$ au$^{-1}$) that borders also the HCM cluster of Athor. 
We make the hypothesis that the group associated to (161) Athor by the HCM is the core of a collisional family and the other asteroids that are within the V-shape and have ($\sin i <$~0.1 and $e >$~0.12) form the family "halo" \citep{Parker2008Icar..198..138P,Broz2013Icar..223..844B} of the Athor family. 

\ca{A fundamental test that these two groups of asteroids are member of the same collisional family, dispersed by the Yarkovsky effect, come from rejecting the null-hypothesis that the asteroids semimajor axis could be derived from a size independent distributions.}
We thus perform the statistical test described in section~\ref{S:StatTest} and we find no cases in $10^6$ trials where we can generate a V-shape like the observed one drawing randomly from the semimajor distribution of all X-type asteroids of the IMB. This demonstrates that the V-shape is not just a consequence of the statistical sampling of an underlying size independent distribution, \ca{but instead it indicates dispersion of asteroids from a common origin due to the Yarkovsky effect, which is a strong indication that they belong to a collisional family}. We use 146 observed bodies, 5 of which are outside the borders of the V-shape. The, average number of simulated bodies that fall outside the V-shape is 23 with a standard deviation of 3.5.

In order to determine the age of the Athor family, as described in section~\ref{S:age}, we need to estimate the drift rate due to the Yarkovsky effect, which depends on a number of parameters such as heliocentric distance, bulk density, rotation period, and thermal inertia. We fix the semimajor axis at the centre of the family ($a_c$ = 2.38 au), we determine the probability distribution for the albedo, rotational period, density, and thermal inertia, which are all the parameters relevant to estimate the value of d$a$/dt in Eq.~\ref{E:InvSlopeAge}. Values of the bolometric Bond's albedo ($A$) are obtained from those of the geometric visible albedo $p_V$ via the equation $A = p_V (0.29 + 0.684 G)$, where $p_V$ and $G$ are given in Tab.~\ref{T:AthorMembers}. It is possible to demonstrate that the distribution of $A$ is well approximated by a Gaussian function centred at $A$ = 0.073 and with $\sigma$ = 0.02. 
None of the family members has known density. We therefore estimate this parameter by taking a weighted mean density of asteroids belonging to the X-complex and with $0.1 \leq p_V \leq 0.3$ (see Tab.~\ref{T:Xdensities}). The weights are given by the inverse square of density uncertainties. We find an average value of 3,500 kg~m$^{-3}$ \ca{and an uncertainty of the mean of 150 kg~m$^{-3}$. We estimate that a value three times larger i.e 450 kg~m$^{-3}$ is more appropriate the the density uncertainty}. For the age determination, we thus assume that the probability function of the density of family members is a Gaussian function centred at \ca{3,500~kg~m$^{-3}$ and with $\sigma =$~450 kg~m$^{-3}$}.
The asteroid (757) Portlandia, with a spherical equivalent diameter of about 33 km, is at the moment of writing the only family member with known thermal inertia value ($\Gamma$) around 60 J~m$^{-2}$~s$^{-0.5}$~K$^{-1}$ \citep{Hanus2018Icar..309..297H}. The other non-family X-complex asteroids with 0.1 $< p_V<$ 0.3 with measured thermal inertia values are (272) Antonia, (413) Edburga, (731) Sorga, (789) Lena, (857) Glasenappia,  and (1013) Tombecka with values of 75, 110, 62, 47, 47, and 55 J~m$^{-2}$~s$^{-0.5}$~K$^{-1}$, which were measured at heliocentric distances ($r_h$) of 2.9, 2.9, 3.3, 2.7,2.3, and 3.1 au, respectively \citep{Hanus2018Icar..309..297H}. Since thermal inertia is temperature dependent \citep{Rozitis2018MNRAS.477.1782R} and the surface temperature of an asteroid is a function of its heliocentric distance, we correct these thermal inertia values to 2.38~au assuming that $\Gamma \propto r_h^{-3/4}$ \citep{Delbo2015aste.book..107D}. We find a mean value of 80 J~m$^{-2}$~s$^{-0.5}$~K$^{-1}$ and a standard deviation of 25 J~m$^{-2}$~s$^{-0.5}$~K$^{-1}$.
We therefore assume a Gaussian with these parameters for  the thermal inertia probability distribution of the family members. 
%
We assume the rotation period of family members to be represented by a uniform distribution with values between 2.41~h and 6.58~h. This is because of the twelve family members with known rotational periods,  (757) Portlandia, (2419) Moldavia, (4353) Onizaki, (5236) Yoko, and (7116) Mentall 
are very close to the inward border of the family V-shape, and thus the best candidates to use for the calculation of d$a$/d$t$ due to the Yarkovsky effect. 
As in the case of the primordial family of \cite{Delbo2017Sci...357.1026M} we find here that the Monte Carlo simulations produce a probability distributions of the ages of the family that is nearly lognormal. The best Gaussian fit to this distribution in $\log T$-space allows us to estimate the most likely age and its formal standard deviations of the family 
to be \ca{3.0$^{+0.5}_{-0.4}$~Gy}. 

\begin{figure}[h!!!]
\includegraphics[width=\columnwidth]{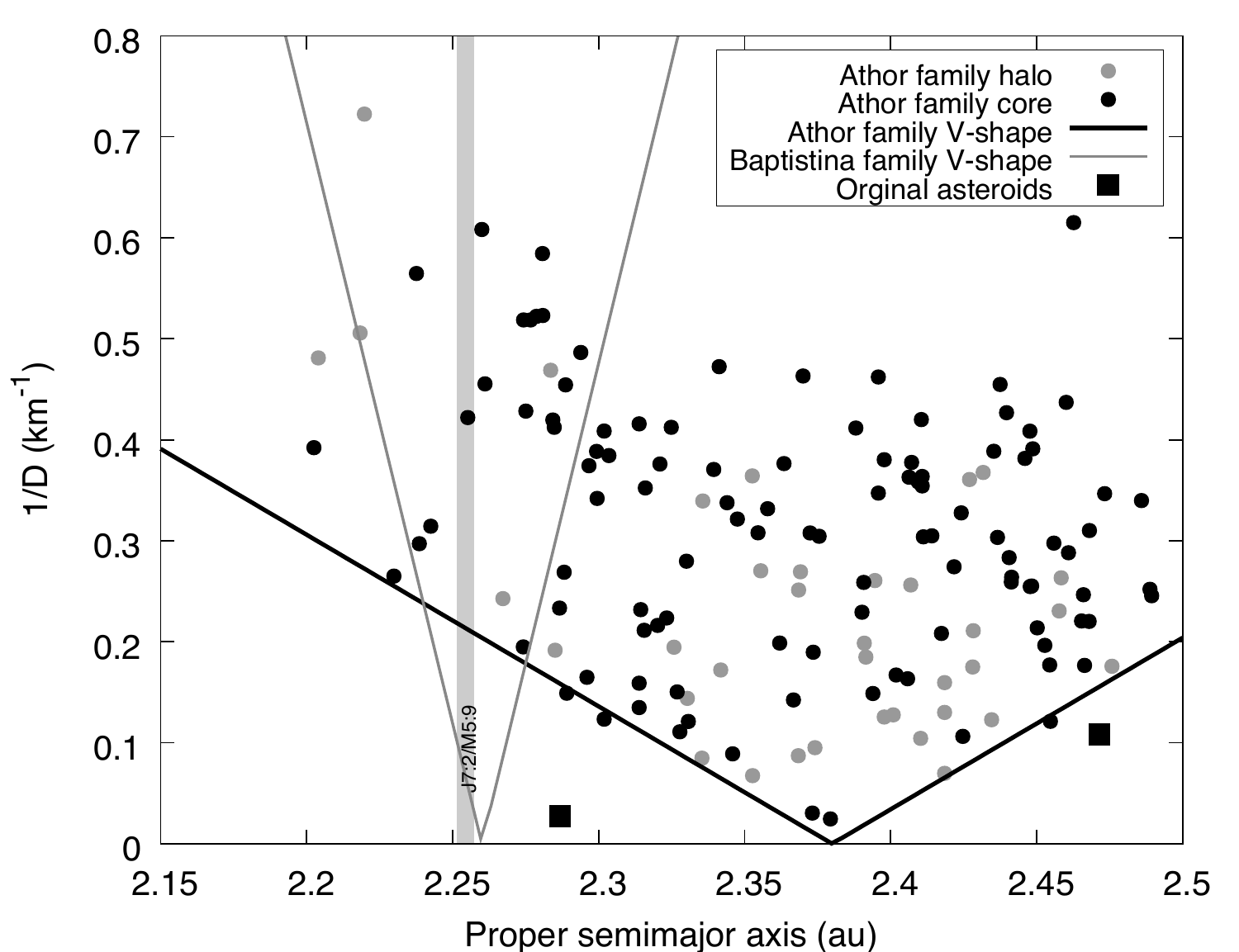}
\includegraphics[width=\columnwidth]{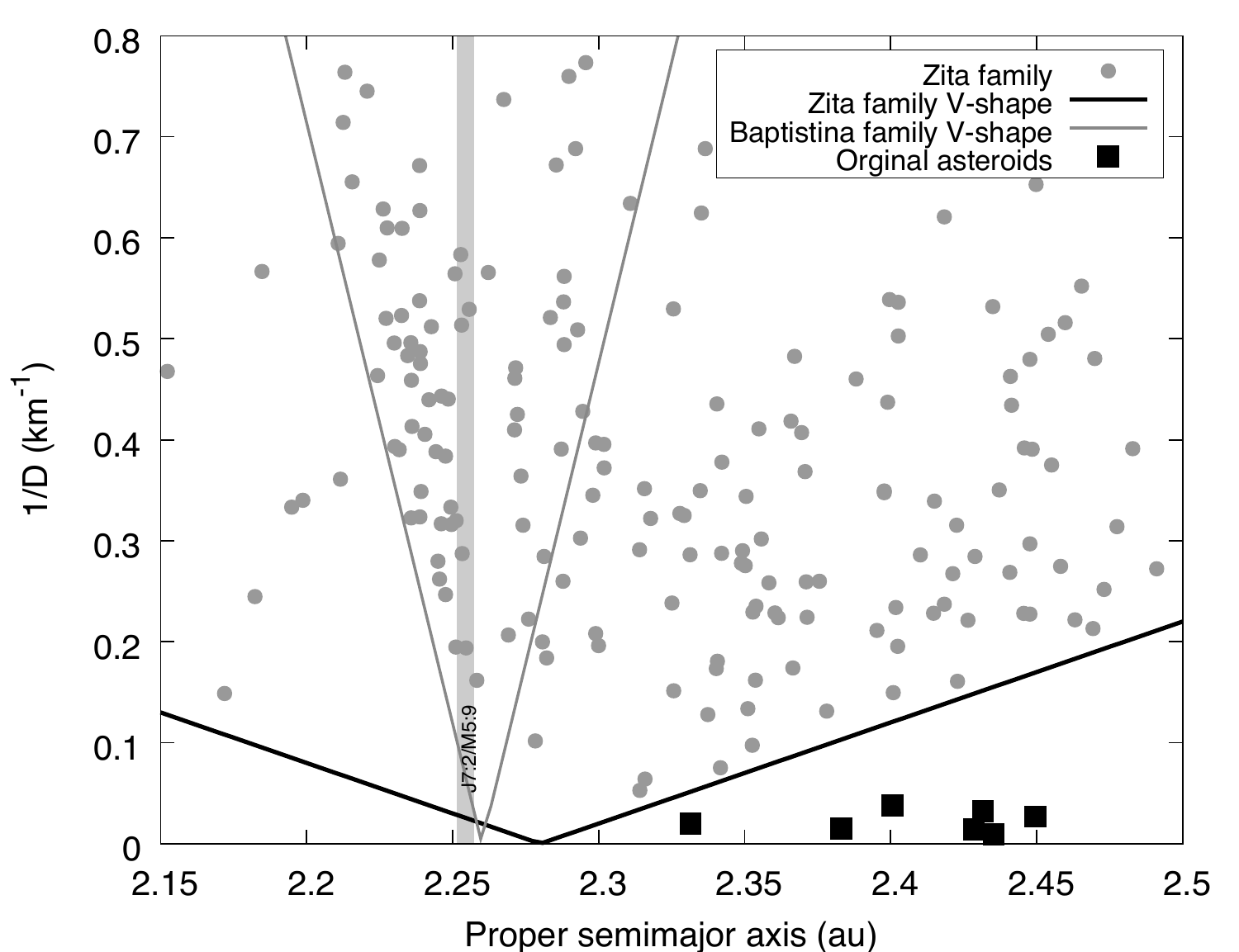}
\caption{\label{F:agapi2Vshape} \ca{Distribution of X-complex asteroids with 0.1~$< p_V <$~0.3 of the inner Main Belt in the ($a,1/D$) space and that do not belong to the Baptistina asteroid family according to \cite{Nesvorny2015aste.book..297N}. Top: asteroids with $\sin i >$ 0.1 and $e <$ 0.12. Objects represented by circles are Athor family members; the squares are planetesimals. Bottom: all the other asteroids that belong neither to Baptistina, nor to the core and halo of Athor. Those represented with circles are Zita family members. Squares are the planetesimals. The position of the overlapping J7:2 MMR and the M5:9 MMR is drawn with a vertical bar. The V-shapes of the Hathor, Zita, and Baptistina families are plotted (but not the members of the Baptistina family). As discussed in section~\ref{S:discussion}, we define as planetesimals those asteroids that cannot be included into V-shapes.}}
\end{figure}

Next, we turn our attention to the ``background'', i.e. those asteroids of the X-type population of the inner Main Belt with 0.1~$< p_V<$~0.3 which remain after we remove the members of the core and halo of the Athor family and the members of the Baptistina family: Fig.~\ref{F:agapi2Vshape} shows that their $(a,1/D)$ distribution appears to have a border that follows the outward slope of the V-shape with parameters $a_c\sim$2.28 au and $K\sim$1 km$^{-1}$ au$^{-1}$, which corresponds to the peak labeled "2" in Fig.~\ref{F:vshapeScan}.

In order to determine the uncertainties on the values of its $a_c$ and $K$ parameters, we perform a Monte Carlo simulation as described in section \ref{S:VshapeMethod}, where the ranges of parameter values explored are $0.0 \leq K \leq 2$ au$^{-1}$~km$^{-1}$ and $2.22 \leq a_c \leq 2.32$ au). We find that the uncertainties on the determination of $a_c$ and  $K$ are 0.02~au and 0.2~au$^{-1}$~km$^{-1}$, respectively. 

We perform the statistical test described in section~\ref{S:StatTest} we find about 2870 cases in $10^6$ trials where we can generate a V-shape as the one observed from a size-independent semimajor axis distribution distribution of asteroids. This demonstrates that also this V-shape is probably not just a consequence of the statistical sampling of an underlying size independent semimajor axis distribution, however with smaller statistical significance than the V-shape of Athor and the V-shape of the primordial family of \cite{Delbo2017Sci...357.1026M}. We use 183 observed bodies, none of which is outside the borders of the V-shape (all the grey points of Fig.~\ref{F:agapi2Vshape}). The average number of simulated bodies that fall outside the V-shape is 4.8 with a standard deviation of 1.8. This means that it is possible that this second V-shape is just due to statistical sampling, despite that this probability is very low ($\sim$0.29\%, which means that this family is robust at 2.98$\sigma$). Assuming similar physical properties as described above the age ratio between this second V-shape and the one centred on (161) Athor is given, to first order, by the ratio of the V-shape slopes, i.e. about 5.16~Gyr. We refine the age, using the same aforementioned Monte Carlo method (taking into account the change of the solar constant with time), we find that the distribution of the family age ($t_0$) values is well represented by a Gaussian function in $\log_{10}(t_0)$ centred at 0.7 and with standard deviation of 0.1. This corresponds to a family age of \ca{5.0$^{+1.6}_{-1.3}$~Gyr}. Given the errors involved in the age estimation process, we interpret that this V-shape can be as old as our Solar System.

The lowest numbered asteroid of this V-shape is the asteroid (689) Zita with a diameter of 15.6~km, while the largest member of this family is the asteroid (1063) Aquilegia, which has a diameter of 19~km. We highlight that the CX classification of Zita comes from multi-filter photometry of \cite{Tholen1989aste.conf..298T}, and its $p_V = 0.10 \pm 0.02$ is also compatible with an object of the C-complex. 

\section{Discussion}
\label{S:discussion}
First of all, we deal with caution the parenthood of the families, because it is possible that neither (161) Athor nor (689) Zita are the parents of their respective families. This is due to the fact that these families are several Gyr-old and have likely lost a significant fraction of their members by collisional and dynamical depletion possibly including their parent bodies or largest remnants \citep[as described by][and references therein]{Delbo2017Sci...357.1026M}. This is particularly important for the Zita family, which we argue that predated the giant planet instability, and to a less extent for the Athor family. \ca{Some of the Zita family members have sizes of a few km. The life time of these bodies in the Main Belt against their breakup by collisions with other asteroids is smaller than the age of the family \citep{Bottke2005Icar..175..111B}. It is therefore likely that they were created by the fragmentation of a larger Zita family member due to a more recent collision than the family forming one. Fragmentation of bodies inside a family produces new asteroids which coordinates in the $(a,1/D)$ space still lay inside the family V-shape.}

Thirteen members of the Athor family have visible spectra from the literature, which we plot in Fig.~\ref{F:athorSpec}. Of these, the asteroid (4353) Onizaki has dubious classification and could be a low-albedo S-type \citep{Bus2002Icar..158..106B}, thus an interloper to this family. The remaining of the objects have spectra similar to the Xc or the Xk average spectra (Fig.~\ref{F:athorSpec}). No family member has near-infrared spectroscopy. Further spectroscopic surveys of members of the Athor family will better constrain the composition of this family (Avdellidou et al., in preparation).  In addition, the  currently on-going space mission Gaia is collecting low resolution spectra of asteroids in the visible \citep{Delbo2012P&SS...73...86D}; the Gaia Data Release 3 in 2021, will contain asteroid spectra and should help to further constrain family membership.

Fourteen core members of the Athor family -- asteroids numbered 2419, 11977, 12425, 17710, 21612, 30141, 30814, 33514, 34545, 42432, 43739, 54169, 68996, 80599 -- and seven members of the halo of Athor family -- 1697 2346 3865 8069 29750 42016 79780 -- were also linked by \cite{Nesvorny2015aste.book..297N} to the family of the asteroid (4) Vesta. None of these asteroids is classified as V-type, which would be expected given the composition of (4) Vesta and its family. 
The geometric visible albedo distribution of these asteroids has a mean value of 0.18 and standard deviation of 0.04, while that of the Vesta family has a mean value of 0.35 and standard deviation of 0.11. The albedo values and spectra or spectrophotometric data of these asteroids are thus incompatible with those of the Vesta family, making them very likely interlopers that are linked to (4) Vesta by the HCM because of their orbital element proximity. 
\ca{While the V-shape family identification is quite robust against including interlopers, in particular because we apply it on compositionally similar asteroids, the presence of interlopers are still possible. In our case, interlopers are mostly the results of uncertainties in the albedo and spectral identification. We estimated the effect of albedo uncertainty on the selection of asteroids following the Monte Carlo method described in section \ref{S:VshapeMethod}. We find the number of asteroids selected by the criterion 0.1~$\leq p_V \leq$~0.3 varies within 6\%. Uncertainty in spectral class assignation adds probably another 5--10\% cents to the number of interlopers.}

In Fig.~\ref{F:agapiAEIvel} one can locate the core of the $\nu_6$ secular resonance at $\sim$2.15 au for $\sin i \sim$~0.1 and Fig.~\ref{F:agapi2Vshape} shows that the inward border of the V-shape of the Athor family crosses $a=$~2.15 au  for $1/D\sim$~0.4~km$^{-1}$. This implies that Athor family can feed the $\nu_6$ with asteroid with $D=2.5$~km or smaller.  However, members of Athor family start to be already removed from the Main Belt at $a\sim$~2.2~au, because of the width of the dynamically unstable region around the $\nu_6$. This indicates that Athor family members with sizes $D \lesssim$~2.5~--~3~km  can leave the Main Belt and become near-Earth asteroids. Another possible escape route from the Main Belt to near-Earth space is the zone of overlapping J7:2 and M5:9 mean motion resonances with Jupiter and Mars, respectively \citep[which is also an important escape route for the Baptistina family members;][]{Bottke2007Natur.449...48B}. Asteroids drifting with $da/dt<0$ from Athor family's centre at $a=$~2.38~au would encounter the overlapping J7:2 and M5:9 mean motion resonance before the $\nu_6$. Indeed, the number density of Athor family members visually appears to decrease inward (to the left) of the overlapping J7:2 and M5:9 (Fig.~\ref{F:agapi2Vshape}). Twelve of the fifteen X-complex NEAs with moderate geometric visible albedo (0.1 $< p_V<$ 0.3) have osculating $\sin i > $~0.1, indicating their likely origin from the \ca{moderate} inclination part of the inner Main Belt, which is also the location of the Athor family. 

Using the aforementioned argument, one can see that the Zita family can deliver asteroids with $1/D\gtrsim$~0.12~km$^{-1}$, i.e. $D \lesssim$~8.3~km to the $\nu_6$ resonance. A visual inspection of Fig.~\ref{F:agapi2Vshape} suggests that the number density of Zita family members firstly increases and then rapidly decreases for $a<$~2.2545~au, where the J7:2 and M5:9 resonances are overlapping. The number density increase is  confined within the inward section (i.e. to the left side) of the V-shape of the Baptistina family (Fig.~\ref{F:agapi2Vshape}) despite that these asteroids are not linked by \cite{Nesvorny2015aste.book..297N} to this family. Out of the 47 asteroids that in Fig.~\ref{F:agapi2Vshape} resides between the inward border of the Baptistina family and the J7:2|M5:9 resonances (i.e. that have $1/D \leq 11.9 \times (a-2.26)$~km$^{-1}$ and $a<$~2.2545~au), 27 are linked to the Flora asteroid family by  \cite{Nesvorny2015aste.book..297N}. We suspect that these asteroids could be in reality Babptistina family members, although are currently linked to the Flora family by the HCM of  \cite{Nesvorny2015aste.book..297N}. 
As in the case of the Athor family, members of the Zita family drifting with $da/dt<0$ from Athor family's centre at $a=$~2.28~au would encounter the overlapping J7:2 and M5:9 MMRs before the $\nu_6$.
Both Athor and Zita families can also deliver asteroids of $D\lesssim$~5~km to the near-Earth space via the 3:1 mean motion resonance with Jupiter. 

While the $e$ and $i$ distributions of the $\sim$3~Gry-old Athor family are still relatively compact, this is not the case for the Zita family. The $e$ and $i$ distributions of the latter family resemble those calculated by \cite{OBrasil2016Icar..266..142B} during the dynamical dispersal of asteroids -- without scattering caused by close enconters with a fifth giant planet -- due to the giant planet instability \citep{Tsiganis2005Natur.435..459T,Morbidelli2015aste.book..493M}. Together with its old age ($>$4~Gry) we interpret the large dispersion in $e$ and $i$, covering the whole phase space of the inner Main Belt, as a sign that the Zita family is primordial, i.e. it formed before the giant planet instability. This instability is needed to explain the orbital structure of the transneptunian objects \citep{Levison2008Icar..196..258L,Nesvorny2018ARA&A..56..137N}, the capture of the Jupiter trojans from objects scattered inward from the primordial transneptunian disk \citep{Morbidelli2005Natur.435..462M,Nesvorny2013ApJ...768...45N},  
the orbital architecture of the asteroid belt \citep{Roig2015AJ....150..186R}, and that of the giant planets irregular satellites \citep{Nesvorny2007AJ....133.1962N}; see \cite{Nesvorny2018ARA&A..56..137N} for a review.
The exact timing of this event is still debatable and matter of current research:  it was initially tied to the Late Heavy bombardment and the formation of the youngest lunar basins i.e. several hundreds ($\sim$700) Myr after the formation of the calcium aluminium rich inclusions \citep{Bouvier2010NatGe...3..637B,Amelin2002Sci...297.1678A}, which is taken as the reference epoch for the beginning of our Solar System.  However,  recent works tend to invoke an earlier instability \citep{Morbidelli2018Icar..305..262M}. 
In addition, simulations by \cite{Nesvorny2018NatAs.tmp..123N} of  the survival of the binary Jupiter Trojan Patroclus--Menoetius,  which was originally embedded in a massive transneptunian disk, but dislodged from this and captured as trojan by Jupiter during the instability, indicate that this disk had to be dispersed within less than 100~Myr after the Solar System origin. Since the dispersion of the transneptunian disk is a consequence of the giant planet instability \citep{Levison2008Icar..196..258L}, one can derive the upper limit of 100~Myr for the time for this event  \citep{Nesvorny2018NatAs.tmp..123N}.  Moreover, \cite{Clement2018Icar..311..340C} propose that the instability occurred only some (1--10) My after the dispersal of the gas of the protoplantery disk.
The age estimate of the Zita family, given its uncertainty, can only constrain the epoch of the giant instability to be $\lesssim$600~Myr after the beginning of the Solar System. 
The major source of uncertainty is the determination of the value of $da/dt$ due to the Yarkovsky effect for family members. It is expected that end-of-mission Gaia data will allow the Yarkosvky $da/dt$ to be constrained for main belt asteroids \citep{Gaia2018A&A...616A..13G}.


Following the logic of \cite{Delbo2017Sci...357.1026M}, we propose that X-complex asteroids with geometric visible albedo between 0.1 and 0.3 of the inner part of the Main Belt have two populations of bodies: one that comprises the bodies inside V-shapes (Athor, Baptistina, and Zita) \ca{but not the parent bodies of the families} and the other one that includes only those 9 bodies that are outside V-shapes \ca{and the parent bodies of the families: i.e. (161) Athor, (298) Baptistina, and (689) Zita}. Objects of the former population are family members and were therefore created as fragments of parent asteroids that broke up during catastrophic collisions. On the other hand, asteroids from the latter population cannot be \ca{created from the fragmentation of the parent body of a family} and, as such, we consider that they formed as planetsimals by dust accretion in the protoplanetary disk. The implication is that the vast majority of the X-types with moderate albedo of the inner Main Belt are genetically related to a few parent bodies -- Zita, Athor, Hertha \citep{Dykhuis2015Icar..252..199D}, and possibly Baptistina \citep{Bottke2007Natur.449...48B}, even though the latter family could be composed by objects which composition is similar to that of impact blackened ordinary chondrite meteorites \citep{Reddy2014Icar..237..116R} and therefore be more similar to S-type asteroids.

\begin{figure*}[ht]
\includegraphics[width=.5\textwidth]{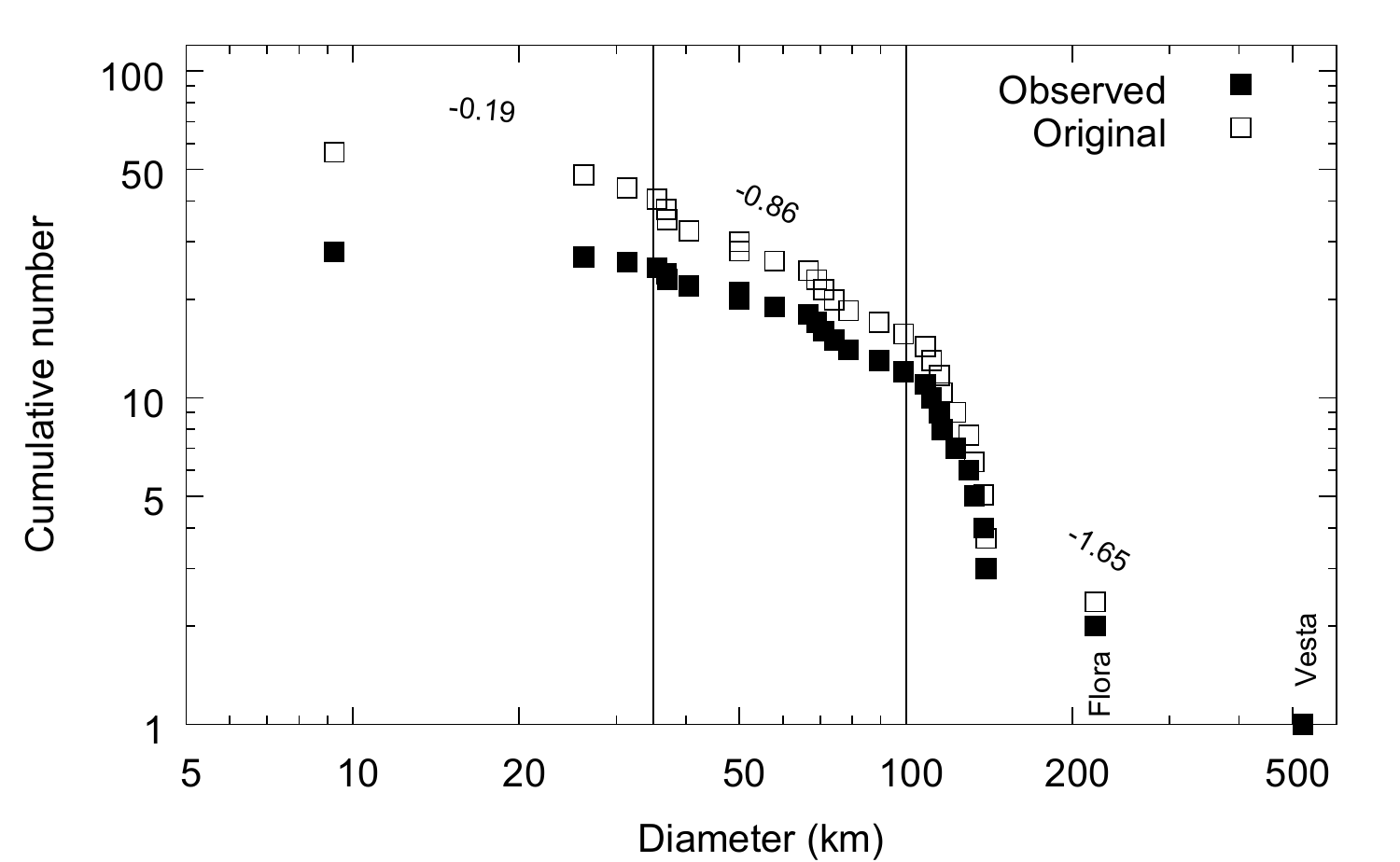}
\includegraphics[width=.5\textwidth]{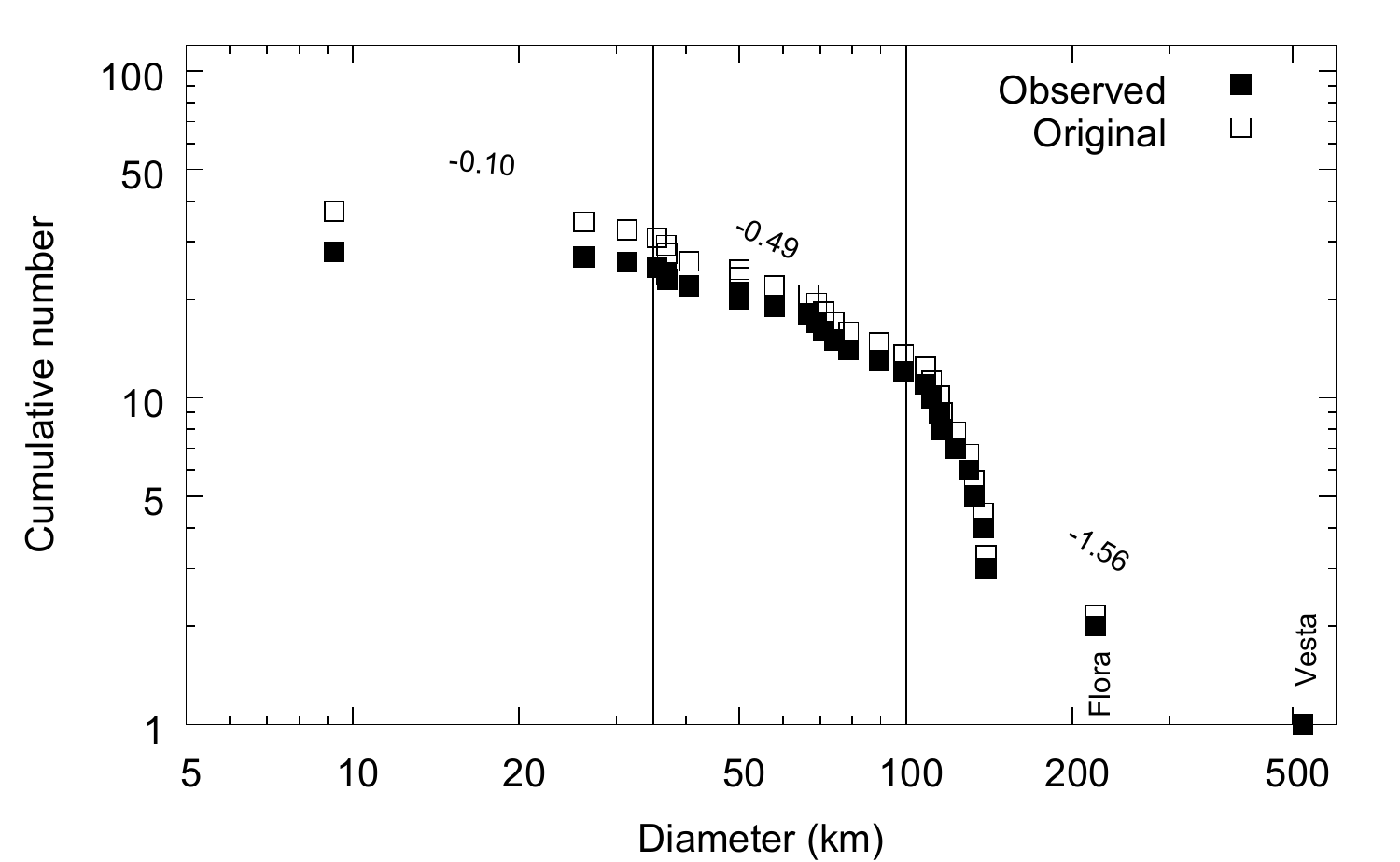}
\caption{\label{F:planetesimalsSFD}Cumulative size distribution of planetesimals. 
\ca{This is the cumulative size distribution of those asteroids that are outside V-shapes and thus probably do not belong to a family. Additionally, in this dataset, are included the parent bodies of families, as originally they belonged to the planetesimal population. The distribution is then corrected for the maximum number of objects that were lost, due to the collisional and dynamical evolution, in order to obtain an upper limit for the distribution of the planetesimals (open squares). }
Functions of the form $N(>D) = N_0 D^{\beta}$, where $N$ is the cumulative number of asteroids, are fitted piecewise in the size ranges $D>$ 100~km, 35 $<D<$ 100~km, and $D<$ 35~km. For the original planetesimals size distribution, we obtain the values of $\beta$ reported by the labels in the plot.  {\it Left:}  the correction to the number of asteroids is calculated taking into account 10~Gyr of collisional evolution in the present Main Belt environment, following the prescription of \cite{Bottke2005Icar..175..111B}.  {\it Right:}  correction to the number of asteroids taking into account 4.5~Gyr of collisional evolution in the present Main Belt environment.}
\end{figure*}

We plot the size distribution of the population of bodies that are not included into V-shapes together with those already identified by \cite{Delbo2017Sci...357.1026M} in Fig.~\ref{F:planetesimalsSFD}. As progenitors of the Athor and Zita families, we also add two asteroids whose effective diameters are 133 and 130~km as deduced by the cubic root of the sum of the cube of the diameters of the members of the Athor (core and halo) and Zita families, respectively. These sizes represent a lower limit to the real sizes of the parents that broke up to form the families.
Following the same procedure of \cite{Delbo2017Sci...357.1026M} we compute an upper limit for the original size distribution of the planetesimals taking into account asteroid dynamical and collisional loss as a stochastic process. The resulting original size distribution (Fig.~\ref{F:planetesimalsSFD}) is still shallower than predicted by some current accretion models \cite{Johansen2015SciA....115109J,Simon2016ApJ...822...55S}, confirming the hypothesis that planetesimals were formed big \cite{Morbidelli2009Icar..204..558M,Bottke2005Icar..175..111B}. Compared to previous results \citep{Delbo2017Sci...357.1026M},  which could not find planetesimals smaller than 35 km in diameter, here we identify planetesimals with $D<$~35 km, but the slope of the planetesimals size distribution below said daimeter is very similar, and even shallower, than the upper bound previously computed \citep[compare Fig.~\ref{F:planetesimalsSFD} with Fig.~4 and Fig.~S7 of] []{Delbo2017Sci...357.1026M}.

 \section{Conclusions} \label{conclusions}
Using our V-shape search method \citep{Bolin2017Icar..282..290B}, we have identified two previously unknown asteroid families in the inner portion of the Main Belt. These families are found amongst the X-type asteroids with moderate geometric visible albedo (0.1$<p_V<$0.3). One of them, the Athor family is found to be $\sim$3~Gyr-old, whereas the second one can be as old as the Solar System. The core of the Athor  family can be found using the HCM and the distributions of the inclinations and eccentricity of the proper orbital elements of its members is relatively compact. On the other hand, in the case of the older family, the eccentricities and inclinations of its member are spread over the entire inner Main Belt. This is an indication that this family could be primordial, which means that it formed before the giant planet instability. 

We show that the vast majority of X-type asteroids of the inner Main Belt can be included into asteroid families. These asteroids are thus genetically related and were generated at different epochs from the catastrophic disruption of very few parent bodies. The X-type asteroids with moderate albedo that are not family members in the inner Main Belt are only nine in number (ten if we include 298 Baptistina). Following the logic of our previous work \citep{Delbo2017Sci...357.1026M}, we can conclude that these bodies were formed by direct accretion of the solids in the protoplanetary disk and are thus surviving planetesimals. We combine the planetesimals found in this work with those from our previous study \citep{Delbo2017Sci...357.1026M} to create a planetesimal size distribution for the inner Main Belt. After correcting the number of objects due to the size-dependent collisional and dynamical evolution of the Main Belt, we find that the distribution is steep for $D>$~100~km but it becomes shallower for $D<$~100~km indicating that $D\sim$~100~km was a preferential size for planetesimal formation.

\begin{acknowledgements} 
The work of C.A. was supported by the French National Research Agency under the project "Investissements d'Avenir" UCA$^{JEDI}$ with the reference number ANR-15-IDEX-01.
M.D. acknowledges support from the French National Program of Planetology (PNP).
This work was also partially supported by the ANR ORIGINS (ANR-18-CE31-0014).
Here we made use of asteroid physical properties data from \url{https://mp3c.oca.eu/}, Observatoire de la C\^ote d'Azur, which database is also mirrored at \url{https://www.cosmos.esa.int/web/astphys}.
We thank Bojan Novakovic for his thorough review. 
\end{acknowledgements}

\bibliographystyle{aa}
\bibliography{mypapers,references,supp} 

\newpage

\begin{appendix}
\section{Supplementary Figures}

\begin{figure*}[]
\includegraphics[width=2\columnwidth]{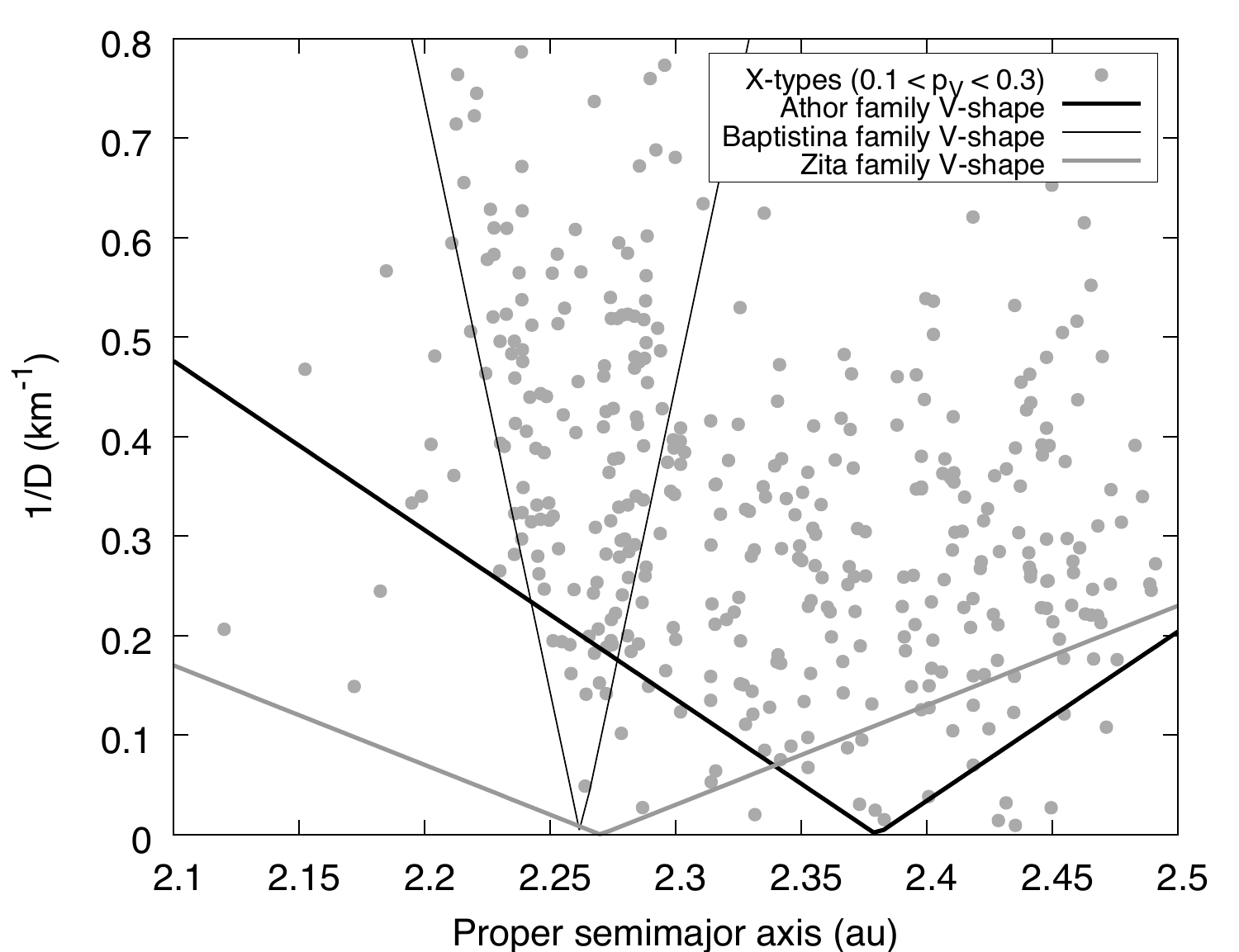}
\caption{\label{F:vshape} V-shapes formed by the X-complex asteroids with geometric visible albedos in the range between 0.1--0.3 of the inner-Main Belt (2.1$<a<$2.5~au). The borders of the V-shapes identified by our method and which position is given in Fig.~\ref{F:vshapeScan} are in one of the panels.}
\end{figure*}

\begin{figure*}[h]
\includegraphics[width=\columnwidth]{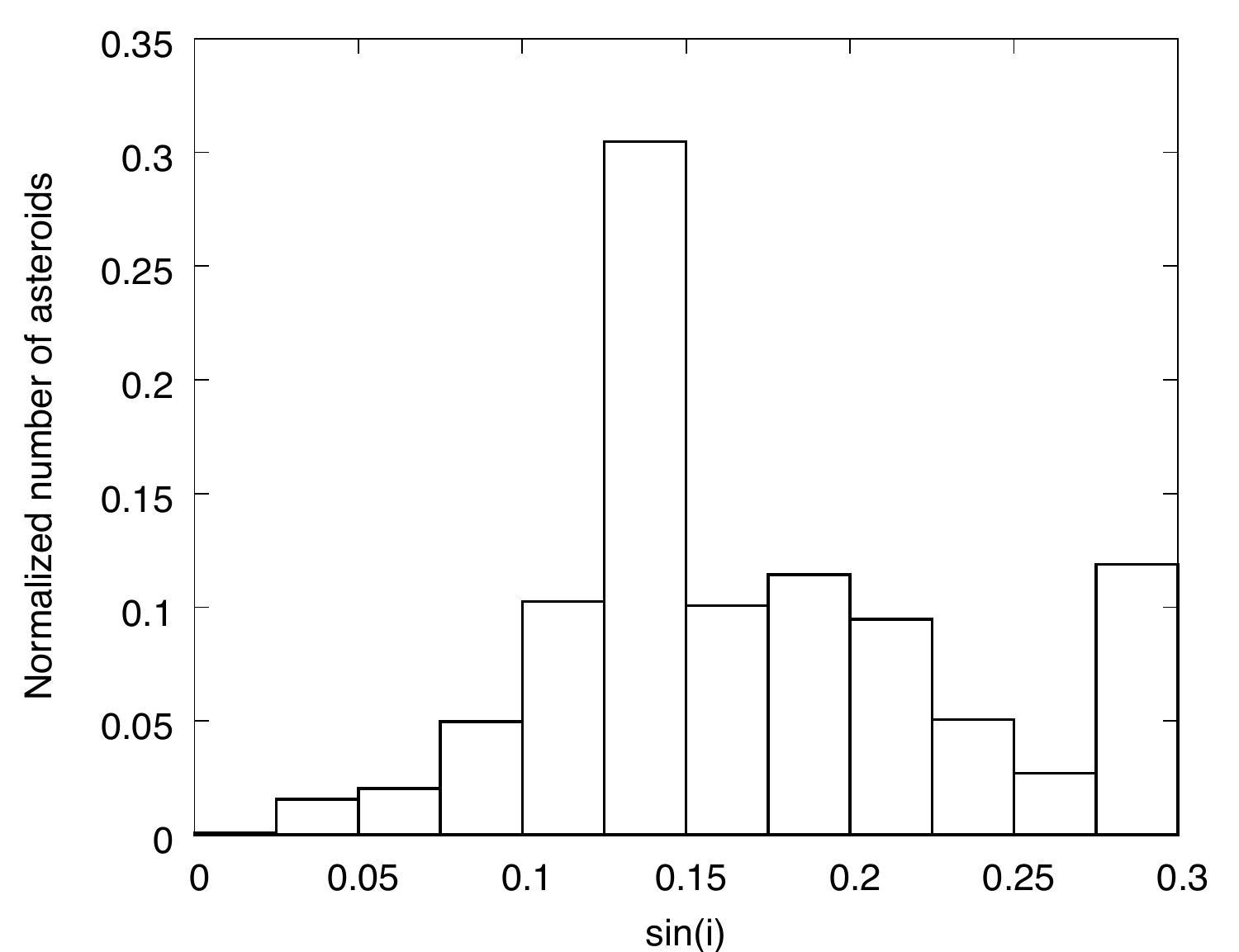}
\includegraphics[width=\columnwidth]{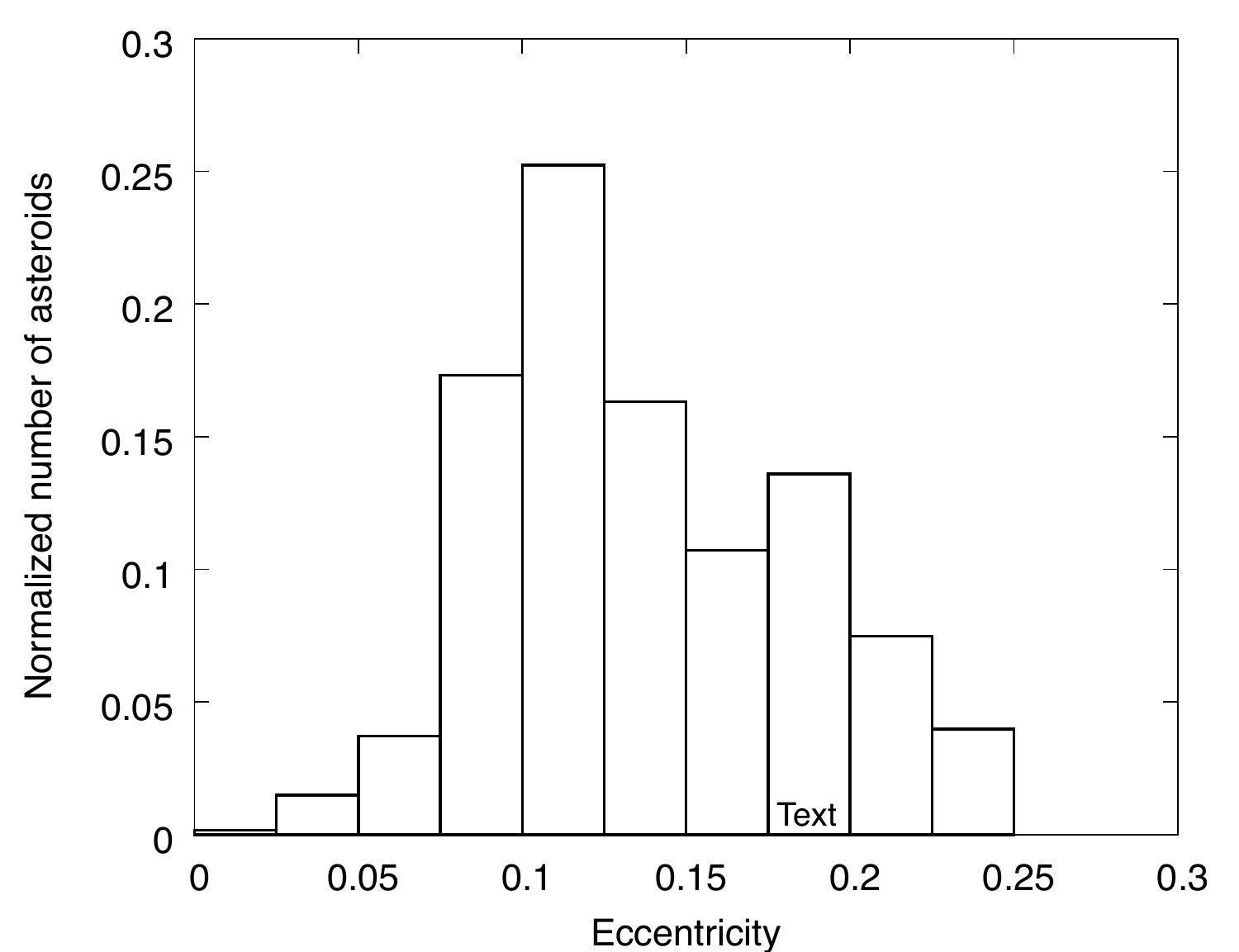}
\caption{\label{F:incEccDistrib} Distribution of the proper eccentricity and sinus of the proper inclination for X-type asteroids inside and near the border of the V-shape with equation $1/D =  1.72~| a-2.38 |$ km$^{-1}$ (see text). The peaks show clustering of these asteroids, indicating the centre of the family.}
\end{figure*}

\begin{figure*}[h]
\centering
\includegraphics[width=2\columnwidth]{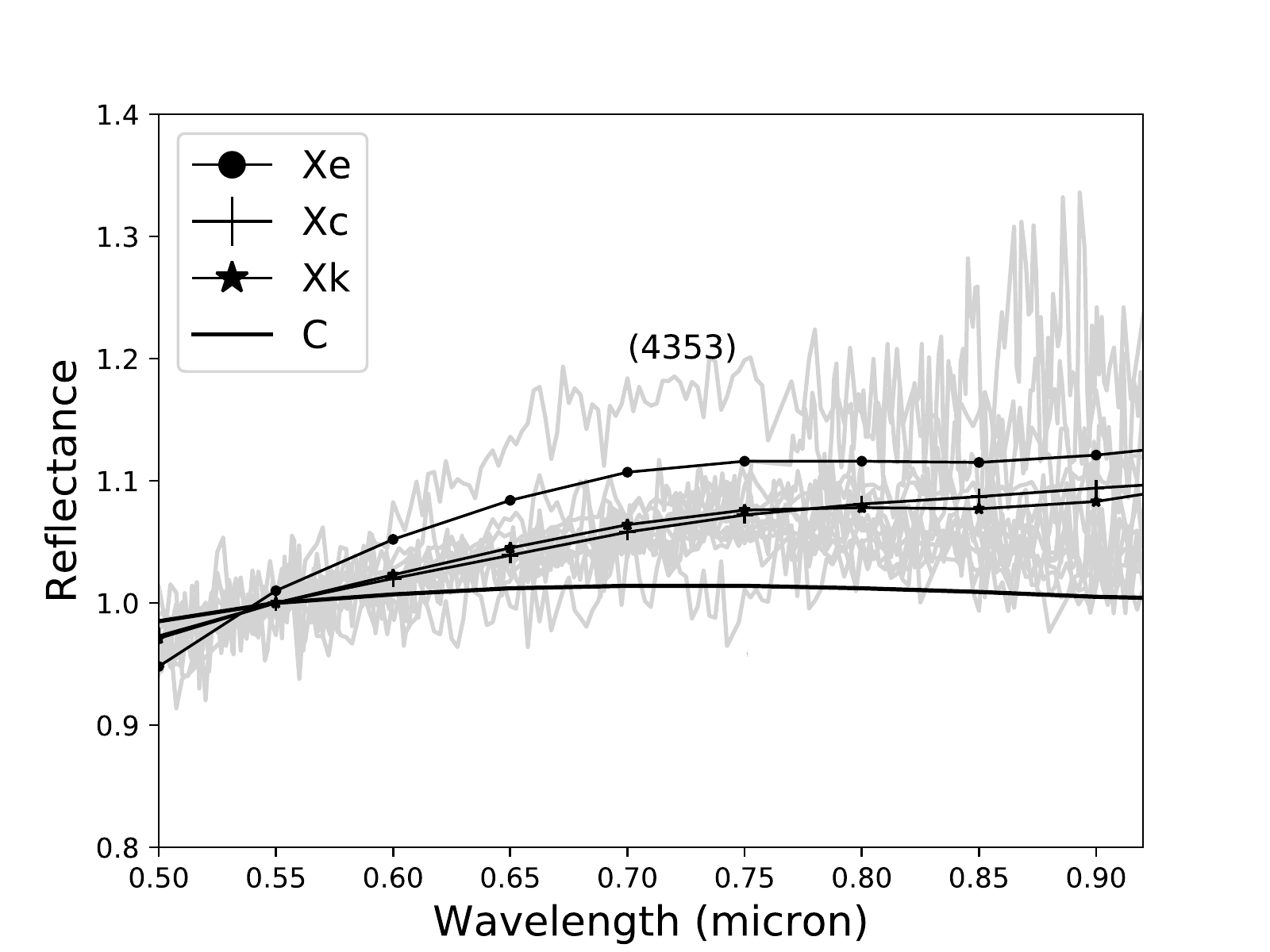}
\caption{\label{F:athorSpec} Reflectance spectra from the literature of Athor family members numbered 161, 757, 1697, 1998, 2194 3007, 3665, 3704 3865, 4353, 4548, 4889, and 4845. The average reflectance spectra of the Xe, Xc, Xk, and C of the \cite{DeMeo2009Icar..202..160D} taxonomy (from \url{http://smass.mit.edu/_documents/busdemeo-meanspectra.xlsx}) are also plotted.
References are: 
(161) from \cite{Bus2002Icar..158..106B},
(757) from \cite{Bus2002Icar..158..106B},
1697 from \cite{Xu1995Icar..115....1X},
(1998) from \cite{Bus2002Icar..158..106B},
(2194) from \cite{Bus2002Icar..158..106B},
(3007) from \cite{Bus2002Icar..158..106B},
3665 from \cite{Xu1995Icar..115....1X},
(3704) from \cite{Bus2002Icar..158..106B},
(3865) from \cite{Bus2002Icar..158..106B},
(4353) from \cite{Bus2002Icar..158..106B},
(4548) from \cite{Bus2002Icar..158..106B},
(4839) from \cite{Bus2002Icar..158..106B},
(4845) from \cite{Bus2002Icar..158..106B}.
}.
\end{figure*}

\section{Supplementary Tables}

 
\onecolumn
\begin{small}
\begin{longtable}{rr|ccccccccccc}
\caption{\label{T:AthorMembers}
Athor family core members.
Numbered references are: 
(1) \cite{DeMeo2009Icar..202..160D}; (2) \cite{Bus2002Icar..158..106B}, (3) \cite{Xu1995Icar..115....1X}, 
(4) \cite{Tholen1989aste.conf..298T}, (5) \cite{Lazzaro2004Icar..172..179L}, (6) \cite{Demeo2013Icar..226..723D}, (7) \cite{Carvano2010A&A...510A..43C}, (8) \cite{Popescu2018A&A...617A..12P},
(32) \cite{Nugent2016AJ....152...63N},
(33) \cite{Nugent2015ApJ...814..117N},
(34) \cite{Masiero2012ApJ...759L...8M},
(35) \cite{Tedesco2002AJ....123.1056T},
(36) \cite{Ryan2010AJ....140..933R},
(37) \cite{Usui2011PASJ...63.1117U},
(38) \cite{Masiero2011ApJ...741...68M},
(39) \cite{Masiero2014ApJ...791..121M}, 
(70) \cite{Mainzer2014ApJ...784..110M},
(71) \cite{Ali-Lagoa2017A&A...603A..55A}.
}\\
\hline\hline
Number & Designation & $a$ & $e$ & sin($i$) & $D$ & $\sigma_D$ & $p_v$ & $\sigma_{p_V}$ & $P$ & Taxonomic & Classi & $D, p_V$\\ 
 &  & (au) & & & (km) & (km) & &  & (hour) & class  & reference & reference\\ 
\hline
   161 &               Athor &  2.3794 & 0.0997 & 0.1535 &   40.840 &  0.520 & 0.233 & 0.007 &   7.28 & Xc,M & 2,4 & 37\\ 
   757 &          Portlandia &  2.3732 & 0.1105 & 0.1357 &   32.890 &  0.240 & 0.220 & 0.020 &   6.58 & Xk,XF,X & 2,4,6 & 39\\ 
  2194 &              Arpola &  2.3278 & 0.0888 & 0.1379 &    9.020 &  0.040 & 0.210 & 0.040 & - & Xc,K,X & 2,6,7 & 39\\ 
  2419 &            Moldavia &  2.2960 & 0.1045 & 0.1149 &    6.070 &  0.150 & 0.240 & 0.030 &   2.41 & C,X & 6,7 & 39\\ 
  4353 &             Onizaki &  2.3459 & 0.1016 & 0.1416 &   11.240 &  0.150 & 0.150 & 0.030 &   4.43 & Xe,X,X,X & 2,3,6,7 & 39\\ 
  5160 &              Camoes &  2.4019 & 0.0606 & 0.1342 &    5.984 &  0.137 & 0.259 & 0.075 & - & X,X,Cgx & 6,7,8 & 38\\ 
  5171 &          Augustesen &  2.4247 & 0.0971 & 0.1328 &    9.410 &  0.330 & 0.108 & 0.008 & 480.00 & K,X & 6,7 & 37\\ 
  5236 &                Yoko &  2.3307 & 0.1028 & 0.1372 &    8.263 &  0.129 & 0.151 & 0.010 &   2.77 & X,S & 7,8 & 38\\ 
  5343 &              Ryzhov &  2.2742 & 0.1206 & 0.1303 &    5.131 &  0.076 & 0.293 & 0.057 & - & X,C & 5,6 & 38\\ 
  6245 &             Ikufumi &  2.3019 & 0.1274 & 0.1512 &    8.105 &  0.148 & 0.129 & 0.025 & - & X & 7 & 38\\ 
  6247 &           Amanogawa &  2.3939 & 0.0433 & 0.1368 &    6.722 &  0.098 & 0.165 & 0.018 &  12.38 & X,X & 6,7 & 38\\ 
  6855 &           Armellini &  2.2891 & 0.0775 & 0.1345 &    6.712 &  0.057 & 0.125 & 0.020 & - & K,X & 6,7 & 38\\ 
  7111 &             1985QA1 &  2.4058 & 0.1150 & 0.1401 &    6.121 &  0.056 & 0.222 & 0.034 &  11.21 & X,X & 6,7 & 38\\ 
  7116 &             Mentall &  2.3139 & 0.1002 & 0.1397 &    7.412 &  0.048 & 0.152 & 0.028 &   6.49 & X,X & 6,7 & 38\\ 
  8381 &           Hauptmann &  2.4664 & 0.1067 & 0.1376 &    5.664 &  0.032 & 0.172 & 0.028 & - & C,X,Kl & 6,7,8 & 38\\ 
 10073 &             1989GJ2 &  2.3620 & 0.0968 & 0.1364 &    5.031 &  0.163 & 0.233 & 0.049 & - & K,X & 6,7 & 38\\ 
 11977 &         Leonrisoldi &  2.3269 & 0.0945 & 0.1188 &    6.650 &  0.190 & 0.145 & 0.022 & - & X,C & 6,7 & 34\\ 
 12425 &             1995VG2 &  2.3156 & 0.0920 & 0.1291 &    4.730 &  0.224 & 0.198 & 0.019 & - & X,C & 6,7 & 38\\ 
 13101 &            Fransson &  2.4609 & 0.0653 & 0.1201 &    3.470 &  0.520 & 0.279 & 0.072 & - & C,CX & 6,7 & 34\\ 
 14317 &             Antonov &  2.4478 & 0.1403 & 0.1163 &    3.928 &  0.323 & 0.262 & 0.115 & - & X,C & 6,7 & 38\\ 
 15705 &              Hautot &  2.4174 & 0.1203 & 0.1381 &    4.803 &  0.102 & 0.160 & 0.026 & - & C,CX & 6,7 & 38\\ 
 15909 &            1997TM17 &  2.2866 & 0.1213 & 0.1279 &    4.286 &  0.109 & 0.208 & 0.011 & - & X,X & 6,7 & 38\\ 
 16007 &         Kaasalainen &  2.3139 & 0.0968 & 0.1375 &    6.291 &  0.048 & 0.108 & 0.020 & - & X,C & 7,8 & 38\\ 
 16258 &           Willhayes &  2.4660 & 0.0942 & 0.1371 &    4.056 &  0.127 & 0.129 & 0.021 & - & C,X & 6,7 & 38\\ 
 17710 &             1997WT2 &  2.2843 & 0.0972 & 0.1190 &    2.383 &  0.345 & 0.163 & 0.048 & - & S,X & 6,7 & 38\\ 
 18578 &              1997XP &  2.2386 & 0.1213 & 0.1360 &    3.366 &  0.059 & 0.179 & 0.027 & - & X,X & 6,7 & 38\\ 
 19207 &             1992QS1 &  2.3667 & 0.1005 & 0.1402 &    7.022 &  0.032 & 0.159 & 0.020 & - & K,X & 6,7 & 38\\ 
 19473 &         Marygardner &  2.4112 & 0.1037 & 0.1340 &    3.290 &  0.441 & 0.236 & 0.032 & - & S,X & 6,7 & 38\\ 
 19564 &          Ajburnetti &  2.3160 & 0.1016 & 0.1316 &    2.838 &  0.204 & 0.166 & 0.034 & - & S,CX & 6,7 & 38\\ 
 20544 &          Kimhansell &  2.3249 & 0.1127 & 0.1470 &    2.425 &  0.540 & 0.228 & 0.119 & - & X & 6 & 38\\ 
 20566 &           Laurielee &  2.3902 & 0.1111 & 0.1427 &    4.363 &  0.050 & 0.185 & 0.012 & - & X,X & 6,7 & 38\\ 
 20652 &           1999TY229 &  2.4545 & 0.1159 & 0.1562 &    5.648 &  0.144 & 0.116 & 0.022 & - & K,CX & 6,7 & 38\\ 
 22761 &             1998YH4 &  2.2299 & 0.0885 & 0.1378 &    3.773 &  0.082 & 0.197 & 0.029 & - & CX,Xt & 7,8 & 38\\ 
 24397 &          Parkerowan &  2.4141 & 0.1099 & 0.1334 &    3.280 &  0.114 & 0.197 & 0.039 & - & X,X,Kl & 6,7,8 & 38\\ 
 26660 &         Samahalpern &  2.3233 & 0.0953 & 0.1451 &    4.473 &  0.166 & 0.184 & 0.017 & - & X,C & 6,7 & 38\\ 
 30141 &           Nelvenzon &  2.3958 & 0.0952 & 0.1262 &    2.880 &  0.614 & 0.162 & 0.069 &  20.74 & X,C & 6,7 & 38\\ 
 30814 &              1990QW &  2.2882 & 0.1021 & 0.1290 &    3.720 &  0.068 & 0.243 & 0.031 & - & C,X & 6,7 & 38\\ 
 33110 &            1998AM10 &  2.4528 & 0.0635 & 0.1288 &    5.092 &  0.163 & 0.247 & 0.050 & - & C,X & 6,7 & 38\\ 
 33486 &             1999GN8 &  2.4406 & 0.1036 & 0.1366 &    3.530 &  0.142 & 0.296 & 0.040 & - & K,X & 6,7 & 38\\ 
 33514 &            1999GF34 &  2.3301 & 0.0996 & 0.1213 &    3.574 &  0.336 & 0.166 & 0.053 & - & X,CX,Kl & 6,7,8 & 38\\ 
 33602 &            1999JW53 &  2.4894 & 0.0870 & 0.1255 &    4.072 &  0.105 & 0.140 & 0.041 & - & C,X & 6,7 & 38\\ 
 34061 &            2000OC48 &  2.3735 & 0.0831 & 0.1475 &    5.274 &  0.205 & 0.160 & 0.036 & - & X,X & 6,7 & 38\\ 
 34297 &           2000QH156 &  2.3724 & 0.1026 & 0.1603 &    3.250 &  0.547 & 0.153 & 0.072 & - & X & 7 & 38\\ 
 34545 &           2000SB234 &  2.4414 & 0.1268 & 0.1114 &    3.789 &  0.140 & 0.148 & 0.021 & - & X & 7 & 38\\ 
 34863 &           2001TP107 &  2.3202 & 0.1024 & 0.1363 &    4.627 &  0.150 & 0.207 & 0.020 & - & C,X,Ad & 6,7,8 & 38\\ 
 35092 &             1990WK6 &  2.3440 & 0.1036 & 0.1315 &    2.961 &  0.622 & 0.221 & 0.129 & - & X,X & 6,7 & 38\\ 
 35739 &            1999GR21 &  2.2425 & 0.0744 & 0.1394 &    3.180 &  0.157 & 0.132 & 0.029 & - & X,C,Cgx & 6,7,8 & 38\\ 
 35753 &            1999GE45 &  2.4413 & 0.1039 & 0.1274 &    3.859 &  0.105 & 0.108 & 0.021 & - & C,X & 6,7 & 38\\ 
 37983 &           1998HB136 &  2.4888 & 0.1193 & 0.1482 &    3.970 &  0.113 & 0.135 & 0.022 & - & X,X & 6,7 & 38\\ 
 38233 &            1999NS57 &  2.3144 & 0.1125 & 0.1428 &    4.313 &  0.024 & 0.115 & 0.013 & - & K,X & 6,7 & 38\\ 
 40694 &           1999RY230 &  2.4680 & 0.1007 & 0.1216 &    4.543 &  0.127 & 0.179 & 0.041 & - & C,X & 6,7 & 38\\ 
 42432 &             1134T-3 &  2.4105 & 0.1036 & 0.1152 &    2.381 &  0.612 & 0.113 & 0.074 & - & X & 7 & 38\\ 
 42633 &            1998FW58 &  2.3546 & 0.0820 & 0.1478 &    3.248 &  1.118 & 0.265 & 0.115 & - & X,X & 6,7 & 38\\ 
 43120 &            1999XB49 &  2.4366 & 0.1002 & 0.1401 &    3.296 &  0.078 & 0.148 & 0.028 & - & X & 7 & 38\\ 
 43656 &           2002ER104 &  2.4217 & 0.1010 & 0.1576 &    3.646 &  0.223 & 0.160 & 0.017 &   3.22 & X,C & 6,7 & 38\\ 
 43739 &             1981EP7 &  2.3755 & 0.1050 & 0.1259 &    3.285 &  0.687 & 0.149 & 0.072 & - & X,X & 6,7 & 38\\ 
 43845 &             1993OS9 &  2.4242 & 0.0980 & 0.1327 &    3.053 &  0.743 & 0.250 & 0.135 & - & C,CX & 6,7 & 38\\ 
 44983 &           1999VU163 &  2.4063 & 0.0963 & 0.1388 &    2.755 &  0.601 & 0.233 & 0.075 & - & CX & 7 & 38\\ 
 54169 &            2000HF57 &  2.3394 & 0.0923 & 0.1288 &    2.698 &  0.301 & 0.243 & 0.063 & - & C,CX & 6,7 & 38\\ 
 54196 &            2000HH78 &  2.4483 & 0.1306 & 0.1354 &    3.919 &  0.169 & 0.138 & 0.028 & - & X & 7 & 38\\ 
 56627 &            2000JF84 &  2.3475 & 0.0649 & 0.1403 &    3.111 &  0.137 & 0.152 & 0.008 & - & X,C & 6,7 & 38\\ 
 56674 &            2000KS77 &  2.4095 & 0.0802 & 0.1423 &    2.791 &  0.943 & 0.189 & 0.079 & - & C,CX & 6,7 & 38\\ 
 58177 &             1990TB6 &  2.3035 & 0.0782 & 0.1233 &    2.602 &  0.459 & 0.238 & 0.078 & - & X & 7 & 38\\ 
 63081 &           2000WP133 &  2.4502 & 0.0630 & 0.1369 &    4.678 &  0.183 & 0.169 & 0.025 & - & CX & 7 & 38\\ 
 65642 &             1981ES2 &  2.2939 & 0.0993 & 0.1372 &    2.056 &  0.553 & 0.264 & 0.101 & - & S,X & 6,7 & 38\\ 
 67449 &           2000QQ166 &  2.4072 & 0.0713 & 0.1552 &    2.648 &  0.553 & 0.230 & 0.175 & - & X & 6 & 38\\ 
 67583 &           2000SZ120 &  2.4559 & 0.1467 & 0.1284 &    3.360 &  0.191 & 0.188 & 0.047 & - & C,X & 6,7 & 38\\ 
 68425 &           2001QS329 &  2.3019 & 0.0767 & 0.1370 &    2.447 &  0.884 & 0.141 & 0.062 & - & X,C & 6,7 & 38\\ 
 68750 &            2002EM68 &  2.2887 & 0.0948 & 0.1499 &    2.201 &  0.078 & 0.100 & 0.004 & - & CX & 7 & 38\\ 
 68996 &           2002TN135 &  2.2993 & 0.0893 & 0.1315 &    2.574 &  0.561 & 0.184 & 0.149 & - & X & 7 & 38\\ 
 70332 &           1999RF166 &  2.3978 & 0.0919 & 0.1454 &    2.630 &  0.437 & 0.147 & 0.079 & - & K,CX & 6,7 & 38\\ 
 70363 &           1999RJ193 &  2.4108 & 0.0194 & 0.1317 &    2.822 &  0.884 & 0.128 & 0.073 & - & K,X & 6,7 & 38\\ 
 70899 &           1999VS176 &  2.4681 & 0.0971 & 0.1359 &    3.224 &  0.517 & 0.246 & 0.042 & - & CX & 7 & 38\\ 
 74119 &            1998QH52 &  2.4547 & 0.0916 & 0.1410 &    8.250 &  1.850 & 0.266 & 0.267 & - & X,X & 6,7 & 34\\ 
 80418 &           1999XA206 &  2.2025 & 0.0908 & 0.1320 &    2.550 &  0.665 & 0.156 & 0.095 & - & C,X & 6,7 & 38\\ 
 80599 &           2000AB148 &  2.2848 & 0.0994 & 0.1256 &    2.425 &  0.767 & 0.208 & 0.164 & - & C,X,Kl & 6,7,8 & 38\\ 
 80835 &             2000DK7 &  2.3908 & 0.1024 & 0.1758 &    3.865 &  0.242 & 0.171 & 0.043 & - & L,X & 6,7 & 38\\ 
 82121 &            2001FQ77 &  2.3139 & 0.0975 & 0.1341 &    2.405 &  0.564 & 0.211 & 0.109 & - & C,X & 6,7 & 38\\ 
 82690 &            2001PB34 &  2.3579 & 0.1007 & 0.1422 &    3.015 &  0.814 & 0.194 & 0.056 & - & C,C,Xt & 6,7,8 & 38\\ 
 85175 &              1990RS &  2.2995 & 0.1200 & 0.1484 &    2.924 &  0.760 & 0.227 & 0.093 & - & C,X & 6,7 & 38\\ 
 85611 &            1998HW16 &  2.4375 & 0.0590 & 0.1398 &    2.199 &  0.579 & 0.175 & 0.093 & - & X,CX & 6,7 & 38\\ 
 88506 &           2001QW140 &  2.3210 & 0.1085 & 0.1353 &    2.659 &  0.671 & 0.228 & 0.144 & - & C,CX & 6,7 & 38\\ 
 92599 &            2000PR19 &  2.3635 & 0.0789 & 0.1351 &    2.656 &  0.266 & 0.158 & 0.023 & - & X,Kl & 7,8 & 38\\ 
 92949 &            2000RS41 &  2.4654 & 0.1007 & 0.1588 &    4.533 &  0.185 & 0.180 & 0.024 & - & X & 6 & 38\\ 
 94674 &            2001XK21 &  2.4487 & 0.0974 & 0.1477 &    2.557 &  0.794 & 0.246 & 0.181 & - & X,X & 6,7 & 38\\ 
 95733 &            2003DG22 &  2.2808 & 0.0972 & 0.1486 &    1.711 &  0.294 & 0.240 & 0.090 & - & X,C & 6,7 & 38\\ 
 95873 &            2003GH37 &  2.3958 & 0.0316 & 0.1403 &    2.164 &  0.177 & 0.137 & 0.044 & - & C,CX & 6,7 & 38\\ 
 98028 &             2000RJ5 &  2.2809 & 0.0856 & 0.1331 &    1.912 &  0.360 & 0.253 & 0.123 & - & X,C & 6,7 & 38\\ 
 98239 &           2000SH156 &  2.2552 & 0.1162 & 0.1432 &    2.370 &  0.695 & 0.125 & 0.030 & - & X,X,Kl & 6,7,8 & 38\\ 
 99490 &           2002CD220 &  2.2743 & 0.1179 & 0.1417 &    1.928 &  0.389 & 0.207 & 0.090 & - & K,X & 6,7 & 38\\ 
101262 &           1998SU104 &  2.3413 & 0.0490 & 0.1415 &    2.117 &  0.224 & 0.273 & 0.085 & - & K,X & 6,7 & 38\\ 
105551 &            2000RW46 &  2.3881 & 0.0550 & 0.1424 &    2.429 &  0.643 & 0.131 & 0.044 & - & X & 7 & 38\\ 
105909 &           2000SZ201 &  2.4108 & 0.1061 & 0.1423 &    2.750 &  0.150 & 0.102 & 0.008 & - & C,X & 6,7 & 38\\ 
108866 &            2001OO99 &  2.4460 & 0.0974 & 0.1496 &    2.621 &  0.495 & 0.135 & 0.066 & - & X,X & 6,7 & 38\\ 
110342 &           2001SN292 &  2.4397 & 0.1060 & 0.1374 &    2.343 &  0.548 & 0.154 & 0.067 & - & X,X & 6,7 & 38\\ 
119194 &           2001QH109 &  2.2611 & 0.0998 & 0.1370 &    2.196 &  0.254 & 0.175 & 0.045 & - & X,C & 6,7 & 38\\ 
119357 &           2001SM238 &  2.4859 & 0.1060 & 0.1584 &    2.943 &  0.792 & 0.141 & 0.059 & - & C,X & 6,7 & 38\\ 
124848 &            2001TZ15 &  2.2787 & 0.1279 & 0.1496 &    1.915 &  0.160 & 0.192 & 0.065 & - & L,X & 6,7 & 38\\ 
125003 &           2001TV154 &  2.4353 & 0.1083 & 0.1246 &    2.573 &  0.445 & 0.184 & 0.050 & - & X & 7 & 38\\ 
125100 &            2001UA34 &  2.2751 & 0.0840 & 0.1351 &    2.334 &  0.351 & 0.155 & 0.053 & - & X,X & 6,7 & 38\\ 
125337 &            2001VE51 &  2.3700 & 0.0685 & 0.1377 &    2.159 &  0.365 & 0.165 & 0.157 & - & X & 7 & 38\\ 
125764 &           2001XG136 &  2.2967 & 0.1071 & 0.1680 &    2.672 &  0.478 & 0.130 & 0.037 & - & C,X & 6,7 & 38\\ 
125895 &           2001XX212 &  2.4477 & 0.0972 & 0.1350 &    2.448 &  0.437 & 0.186 & 0.108 & - & X & 7 & 38\\ 
126771 &            2002DU15 &  2.4733 & 0.1325 & 0.1341 &    2.885 &  0.378 & 0.193 & 0.080 & - & CX & 7 & 38\\ 
132277 &           2002EV150 &  2.4601 & 0.1192 & 0.1372 &    2.288 &  0.586 & 0.194 & 0.087 & - & X & 7 & 38\\ 
145964 &             1999YY7 &  2.2376 & 0.1156 & 0.1317 &    1.771 &  0.163 & 0.246 & 0.021 & - & X,C & 6,7 & 38\\ 
208206 &            2000ST24 &  2.2600 & 0.1011 & 0.1455 &    1.644 &  0.296 & 0.216 & 0.044 & - & K,X & 6,7 & 38\\ 
232377 &            2003BN13 &  2.2767 & 0.1117 & 0.1306 &    1.928 &  0.319 & 0.109 & 0.049 & - & X,CX & 6,7 & 38\\ 
243637 &           Frosinone &  2.3785 & 0.0771 & 0.1339 &    1.508 &  0.313 & 0.195 & 0.079 & - & CX & 7 & 38\\ 
264495 &            2001QB10 &  2.4627 & 0.1168 & 0.1381 &    1.626 &  0.378 & 0.202 & 0.102 & - & CX & 7 & 38\\ 

\hline
\hline
\end{longtable}

\newpage
\begin{longtable}{rr|ccccccccccc}
\caption{\label{table2}
Athor family halo members. For references see Tab~\ref{T:AthorMembers}.}\\
\hline\hline
Number & Designation & $a$ & $e$ & sin($i$) & $D$ & $\sigma_D$ & $p_v$ & $\sigma_{p_V}$ & $P$ & Taxonomic & Class & $D, p_V$\\ 
 &  & (au) & & & (km) & (km) & &  & (hour) & class  & reference & reference\\ 
\hline
  1490 &             Limpopo &  2.3527 & 0.0951 & 0.1911 &   14.840 &  0.110 & 0.170 & 0.030 &   6.65 & Xc & 2 & 39\\ 
  1697 &         Koskenniemi &  2.3741 & 0.1145 & 0.1057 &   10.520 &  0.660 & 0.150 & 0.021 & - & TX & 3 & 37\\ 
  1998 &              Titius &  2.4185 & 0.0848 & 0.1355 &   14.340 &  0.220 & 0.100 & 0.040 &   6.13 & Xc,C,C & 2,6,7 & 39\\ 
  2015 &        Kachuevskaya &  2.3354 & 0.1023 & 0.2098 &   11.810 &  0.150 & 0.140 & 0.020 &  42.53 & X,C & 6,7 & 39\\ 
  3007 &              Reaves &  2.3684 & 0.0921 & 0.1321 &   11.468 &  0.052 & 0.147 & 0.022 &   4.16 & X & 2 & 38\\ 
  3665 &          Fitzgerald &  2.4185 & 0.1112 & 0.2520 &    7.690 &  1.770 & 0.200 & 0.120 &   2.41 & X,C,X & 3,6,7 & 33\\ 
  3704 &            Gaoshiqi &  2.4103 & 0.0552 & 0.1149 &    9.580 &  0.070 & 0.210 & 0.040 &   9.77 & Xk & 2 & 39\\ 
  3865 &           Lindbloom &  2.3978 & 0.0938 & 0.1230 &    7.972 &  0.068 & 0.181 & 0.019 & - & Xc & 2 & 38\\ 
  4548 &              Wielen &  2.2851 & 0.0965 & 0.1354 &    5.220 &  0.090 & 0.179 & 0.030 & - & Xc,S & 2,6 & 38\\ 
  4760 &           Jia-xiang &  2.3258 & 0.0785 & 0.1754 &    5.137 &  0.036 & 0.228 & 0.042 & - & X,X & 6,7 & 38\\ 
  4839 &         Daisetsuzan &  2.4346 & 0.0369 & 0.1328 &    8.141 &  0.099 & 0.202 & 0.032 & - & Xc & 2 & 38\\ 
  4845 &            Tsubetsu &  2.4009 & 0.0825 & 0.1248 &    7.838 &  0.052 & 0.182 & 0.027 & - & X & 2 & 38\\ 
  6123 &         Aristoteles &  2.3304 & 0.0276 & 0.1678 &    6.946 &  0.066 & 0.164 & 0.020 & - & X,X & 6,7 & 38\\ 
  7447 &      Marcusaurelius &  2.3910 & 0.0751 & 0.1270 &    5.034 &  0.122 & 0.220 & 0.010 & - & Xt & 8 & 38\\ 
  8069 &            Benweiss &  2.3418 & 0.1060 & 0.1057 &    5.812 &  0.037 & 0.185 & 0.047 & - & L,DL,Xt & 6,7,8 & 38\\ 
 10368 &              Kozuki &  2.3915 & 0.0814 & 0.1727 &    5.412 &  0.063 & 0.200 & 0.055 & - & X,X & 6,7 & 38\\ 
 13274 &            Roygross &  2.4584 & 0.0354 & 0.1393 &    3.798 &  0.171 & 0.134 & 0.017 & - & CX & 7 & 38\\ 
 20508 &            1999RL25 &  2.4758 & 0.0652 & 0.2791 &    5.689 &  0.113 & 0.198 & 0.022 & - & X,X & 6,7 & 38\\ 
 24053 &          Shinichiro &  2.2183 & 0.0737 & 0.1154 &    1.977 &  0.343 & 0.216 & 0.065 &   6.21 & X & 7 & 38\\ 
 29750 &           Chleborad &  2.4281 & 0.0836 & 0.1103 &    5.714 &  0.062 & 0.155 & 0.010 & - & Xt & 8 & 38\\ 
 33616 &            1999JR64 &  2.4185 & 0.1163 & 0.2469 &    6.269 &  0.125 & 0.149 & 0.021 & - & X & 7 & 38\\ 
 36398 &            2000OQ45 &  2.4283 & 0.0908 & 0.2002 &    4.740 &  0.120 & 0.198 & 0.019 & - & X,C & 6,7 & 38\\ 
 42016 &            2000YZ68 &  2.3946 & 0.1178 & 0.1071 &    3.838 &  0.143 & 0.120 & 0.012 & - & X,X & 6,7 & 38\\ 
 42920 &             1999SA8 &  2.2672 & 0.1191 & 0.1027 &    4.120 &  0.950 & 0.130 & 0.070 &   4.90 & CX & 7 & 32\\ 
 49869 &           1999XG115 &  2.2040 & 0.0825 & 0.1100 &    2.079 &  0.385 & 0.178 & 0.106 & - & CX & 7 & 38\\ 
 61224 &            2000OO13 &  2.3685 & 0.0396 & 0.1125 &    3.980 &  0.129 & 0.161 & 0.034 & - & Xt & 8 & 38\\ 
 67993 &            2000XD24 &  2.3526 & 0.0117 & 0.1577 &    2.746 &  0.778 & 0.195 & 0.149 & - & X & 7 & 38\\ 
 71788 &           2000ST180 &  2.2836 & 0.0420 & 0.1423 &    2.133 &  0.615 & 0.245 & 0.197 & - & C,X & 6,7 & 38\\ 
 72799 &           2001FC188 &  2.4577 & 0.0317 & 0.2220 &    4.339 &  0.218 & 0.113 & 0.016 & - & S,X,S & 6,7,8 & 38\\ 
 79780 &            1998US37 &  2.3357 & 0.1181 & 0.1114 &    2.946 &  0.642 & 0.154 & 0.052 & - & X,Ad & 7,8 & 38\\ 
102416 &           1999TL184 &  2.3555 & 0.0237 & 0.2298 &    3.700 &  0.110 & 0.107 & 0.023 & - & L,Xt & 7,8 & 34\\ 
102808 &           1999VG172 &  2.4317 & 0.0825 & 0.1764 &    2.720 &  0.629 & 0.137 & 0.093 & - & C,X & 6,7 & 38\\ 
116046 &           2003WZ102 &  2.4270 & 0.0841 & 0.2495 &    2.773 &  0.122 & 0.252 & 0.054 & - & Xt & 8 & 38\\ 
122409 &            2000QE89 &  2.4069 & 0.0631 & 0.1829 &    3.903 &  0.155 & 0.168 & 0.022 & - & X,X & 6,7 & 38\\ 
184012 &            2004FT12 &  2.3691 & 0.0783 & 0.2386 &    3.714 &  0.176 & 0.106 & 0.016 & - & CX & 7 & 38\\ 
243735 &             2000OT1 &  2.2198 & 0.0833 & 0.1605 &    1.384 &  0.237 & 0.211 & 0.091 & - & X & 7 & 38\\ 

\hline
\hline
\end{longtable}

\newpage
\begin{longtable}{rr|ccccccccccc}
\caption{\label{table2}
Zita family members.
For references see Tab~\ref{T:AthorMembers}.}\\
\hline\hline
Number & Designation & $a$ & $e$ & sin($i$) & $D$ & $\sigma_D$ & $p_v$ & $\sigma_{p_V}$ & $P$ & Taxonomic & Class & $D, p_V$\\ 
 &  & (au) & & & (km) & (km) & &  & (hour) & class  & reference & reference\\ 
\hline
   689 &                Zita &  2.3159 & 0.1878 & 0.0956 &   15.620 &  0.230 & 0.100 & 0.020 &   6.43 & CX & 4 & 39\\ 
  1063 &           Aquilegia &  2.3141 & 0.0795 & 0.0937 &   18.930 &  0.370 & 0.139 & 0.006 &   5.79 & X,K,X & 3,6,7 & 37\\ 
  2233 &           Kuznetsov &  2.2783 & 0.0775 & 0.0719 &    9.830 &  0.660 & 0.153 & 0.021 &   5.03 & K,X & 6,7 & 37\\ 
  2328 &             Robeson &  2.3417 & 0.2007 & 0.1749 &   13.300 &  0.460 & 0.105 & 0.008 &  18.63 & C,C,X & 2,6,7 & 37\\ 
  2968 &               Iliya &  2.3714 & 0.2729 & 0.1689 &    4.460 &  0.960 & 0.130 & 0.080 & - & C,X,Cgx & 6,7,8 & 33\\ 
  2991 &               Bilbo &  2.3374 & 0.1968 & 0.0704 &    7.818 &  0.147 & 0.115 & 0.013 &   4.06 & Ch,X,C & 5,6,7 & 38\\ 
  3137 &               Horky &  2.4009 & 0.1835 & 0.0589 &    6.685 &  0.089 & 0.207 & 0.030 & - & C,X,CX & 2,6,7 & 38\\ 
  3375 &                 Amy &  2.1719 & 0.0909 & 0.0290 &    6.721 &  0.118 & 0.124 & 0.005 &   3.26 & C,X,C & 2,6,7 & 38\\ 
  3427 &        Szentmartoni &  2.2808 & 0.1080 & 0.0552 &    5.003 &  0.070 & 0.224 & 0.034 & - & Xt & 8 & 38\\ 
  3461 &         Mandelshtam &  2.3781 & 0.1170 & 0.0423 &    7.615 &  0.097 & 0.160 & 0.017 &   2.85 & X,XL & 6,7 & 38\\ 
  4095 &        Ishizuchisan &  2.1201 & 0.0631 & 0.0546 &    4.845 &  0.204 & 0.189 & 0.027 & - & S,S,Xt & 6,7,8 & 38\\ 
  4256 &          Kagamigawa &  2.3511 & 0.0516 & 0.0393 &    7.480 &  0.070 & 0.230 & 0.020 & - & Xc & 2 & 39\\ 
  4838 &      Billmclaughlin &  2.3526 & 0.1461 & 0.1628 &   10.250 &  0.190 & 0.140 & 0.010 &   5.20 & Xc & 2 & 39\\ 
  6182 &            Katygord &  2.2546 & 0.1822 & 0.1006 &    5.158 &  0.051 & 0.185 & 0.024 & - & X,X & 6,7 & 38\\ 
  7739 &                Cech &  2.3251 & 0.1682 & 0.0567 &    4.196 &  0.091 & 0.174 & 0.046 & - & X,X & 6,7 & 38\\ 
  9388 &              Takeno &  2.2476 & 0.1599 & 0.0684 &    4.053 &  0.345 & 0.129 & 0.013 & - & C,CX,Xt & 6,7,8 & 38\\ 
  9602 &                 Oya &  2.2822 & 0.1166 & 0.0390 &    5.436 &  0.076 & 0.165 & 0.024 & - & X,X & 6,7 & 38\\ 
  9735 &              1986JD &  2.3537 & 0.1694 & 0.1383 &    6.172 &  0.085 & 0.168 & 0.040 & - & S,X & 6,7 & 38\\ 
 10309 &             1990QC6 &  2.3278 & 0.1702 & 0.0046 &    3.058 &  0.428 & 0.273 & 0.077 & - & Xt & 8 & 38\\ 
 10359 &            1993TU36 &  2.4229 & 0.1291 & 0.1672 &    6.217 &  0.165 & 0.263 & 0.038 &   2.41 & X & 7 & 38\\ 
 10991 &               Dulov &  2.3539 & 0.1917 & 0.0530 &    4.252 &  0.028 & 0.162 & 0.022 & - & C,X & 6,7 & 38\\ 
 11286 &             1990RO8 &  2.3756 & 0.1657 & 0.0540 &    3.848 &  0.050 & 0.191 & 0.014 &  14.81 & C,X & 6,7 & 38\\ 
 12177 &             Raharto &  2.3978 & 0.1194 & 0.0434 &    2.864 &  0.436 & 0.236 & 0.043 & - & X,XD & 6,7 & 38\\ 
 12582 &            1999RY34 &  2.2511 & 0.1612 & 0.0346 &    5.134 &  0.147 & 0.243 & 0.059 & - & Xt & 8 & 38\\ 
 14026 &            Esquerdo &  2.3604 & 0.1451 & 0.0437 &    4.377 &  0.194 & 0.111 & 0.010 & - & X,C & 6,7 & 38\\ 
 14413 &              Geiger &  2.2878 & 0.1348 & 0.0976 &    3.848 &  0.274 & 0.227 & 0.054 & - & C,CX & 6,7 & 38\\ 
 14505 &           Barentine &  2.3583 & 0.2040 & 0.0428 &    3.870 &  0.186 & 0.155 & 0.021 &   2.89 & X,L & 6,7 & 38\\ 
 14950 &             1996BE2 &  2.2583 & 0.1504 & 0.1022 &    6.180 &  0.050 & 0.190 & 0.030 &   3.28 & C,X & 6,7 & 39\\ 
 15793 &            1993TG19 &  2.4102 & 0.1862 & 0.0637 &    3.497 &  0.926 & 0.132 & 0.092 &  22.40 & CX & 7 & 38\\ 
 15996 &            1998YC12 &  2.3000 & 0.1772 & 0.1398 &    5.099 &  0.157 & 0.142 & 0.025 & - & X,X & 6,7 & 38\\ 
 16177 &              Pelzer &  2.3711 & 0.1389 & 0.0422 &    3.857 &  0.262 & 0.157 & 0.024 & - & S,X & 6,7 & 38\\ 
 16273 &              Oneill &  2.2990 & 0.0905 & 0.0765 &    4.807 &  0.049 & 0.133 & 0.017 & - & X,D & 6,7 & 38\\ 
 16821 &             1997VZ4 &  2.2691 & 0.1483 & 0.1011 &    4.838 &  0.083 & 0.171 & 0.016 & - & C,X & 6,7 & 38\\ 
 16967 &         Marcosbosso &  2.2938 & 0.1211 & 0.0628 &    3.304 &  0.101 & 0.194 & 0.031 & - & X,C & 6,7 & 38\\ 
 17635 &             1996OC1 &  2.4024 & 0.1199 & 0.0997 &    5.117 &  0.166 & 0.107 & 0.024 & - & X & 6 & 38\\ 
 18866 &           1999RA208 &  2.4265 & 0.1741 & 0.2095 &    4.520 &  0.128 & 0.286 & 0.036 & - & C,X,Kl & 6,7,8 & 38\\ 
 19531 &             Charton &  2.2496 & 0.1421 & 0.1091 &    3.164 &  0.130 & 0.101 & 0.013 & - & X & 7 & 38\\ 
 19633 &              Rusjan &  2.4456 & 0.2105 & 0.2211 &    4.385 &  0.297 & 0.277 & 0.096 & - & Xt & 8 & 38\\ 
 20458 &            1999LZ21 &  2.3421 & 0.1780 & 0.1160 &    3.478 &  0.167 & 0.111 & 0.017 & - & X & 7 & 38\\ 
 21091 &             1992DK8 &  2.2455 & 0.1381 & 0.1034 &    3.816 &  0.043 & 0.138 & 0.014 & - & K,CX & 6,7 & 38\\ 
 21475 &          Jasonclain &  2.1823 & 0.0648 & 0.0419 &    4.086 &  0.144 & 0.127 & 0.027 & - & X,D & 6,7 & 38\\ 
 22459 &             1997AD2 &  2.3257 & 0.0740 & 0.0808 &    6.600 &  1.000 & 0.100 & 0.030 & - & X,CX & 6,7 & 71\\ 
 22571 &         Letianzhang &  2.2388 & 0.1536 & 0.0955 &    3.090 &  0.670 & 0.120 & 0.060 & - & X,C & 6,7 & 33\\ 
 23996 &            1999RT27 &  2.3707 & 0.1705 & 0.0505 &    2.714 &  0.398 & 0.182 & 0.066 & - & Xt & 8 & 38\\ 
 25037 &            1998QC37 &  2.1987 & 0.2186 & 0.0884 &    2.940 &  0.670 & 0.210 & 0.090 & - & C,X,Ds & 6,7,8 & 33\\ 
 26014 &             2051P-L &  2.4485 & 0.1760 & 0.0597 &    2.560 &  0.340 & 0.186 & 0.052 & - & X,CX & 6,7 & 34\\ 
 26082 &            1981EB11 &  2.4458 & 0.1758 & 0.0623 &    2.552 &  0.497 & 0.226 & 0.101 & - & C,X & 6,7 & 38\\ 
 26226 &             1998GJ1 &  2.3615 & 0.0407 & 0.0777 &    4.466 &  0.140 & 0.185 & 0.047 & - & X & 7 & 38\\ 
 27043 &            1998RS71 &  2.1949 & 0.0758 & 0.0970 &    3.000 &  0.200 & 0.196 & 0.021 & - & X & 7 & 38\\ 
 27740 &         Obatomoyuki &  2.3953 & 0.1742 & 0.1792 &    4.736 &  0.195 & 0.261 & 0.055 & - & X,C & 6,7 & 38\\ 
 28046 &            1998HB14 &  2.3407 & 0.1526 & 0.1922 &    5.536 &  0.348 & 0.174 & 0.029 & - & X,X & 6,7 & 38\\ 
 28805 &            2000HY85 &  2.4289 & 0.1581 & 0.0916 &    3.515 &  0.130 & 0.299 & 0.041 & - & C,X & 6,7 & 38\\ 
 29442 &             1997NS4 &  2.3978 & 0.1357 & 0.1631 &    2.877 &  0.531 & 0.234 & 0.084 & - & C,CX & 6,7 & 38\\ 
 29652 &             1998WD9 &  2.3313 & 0.1423 & 0.1524 &    3.494 &  0.244 & 0.145 & 0.031 & - & CX & 7 & 38\\ 
 29660 &       Jessmacalpine &  2.2928 & 0.1680 & 0.0565 &    1.965 &  0.442 & 0.240 & 0.091 & - & X,C & 6,7 & 38\\ 
 30244 &            Linhpham &  2.2116 & 0.1774 & 0.1144 &    2.770 &  0.820 & 0.130 & 0.090 & - & X,C,S & 6,7,8 & 32\\ 
 30777 &             1987SB3 &  2.2450 & 0.1348 & 0.0945 &    3.574 &  0.290 & 0.152 & 0.036 & - & X,C & 6,7 & 38\\ 
 30819 &             1990RL2 &  2.3403 & 0.1873 & 0.1741 &    5.763 &  0.037 & 0.101 & 0.023 & - & C,X & 6,7 & 38\\ 
 31110 &              Clapas &  2.2716 & 0.1772 & 0.0671 &    2.122 &  0.068 & 0.171 & 0.031 & - & S,X & 6,7 & 38\\ 
 31682 &              Kinsey &  2.4213 & 0.0864 & 0.0917 &    3.740 &  0.194 & 0.200 & 0.048 & - & C,X & 6,7 & 38\\ 
 31902 &         Raymondwang &  2.2243 & 0.1394 & 0.0534 &    2.157 &  0.374 & 0.263 & 0.088 & - & S,S,Xt & 6,7,8 & 38\\ 
 31937 &          Kangsunwoo &  2.2302 & 0.1516 & 0.0894 &    2.542 &  0.005 & 0.191 & 0.060 & - & C,CX & 6,7 & 38\\ 
 32079 &      Hughsavoldelli &  2.2476 & 0.1329 & 0.1233 &    2.605 &  0.033 & 0.223 & 0.020 & - & K,CX & 6,7 & 38\\ 
 33234 &             1998GL7 &  2.3665 & 0.1877 & 0.2377 &    5.748 &  0.198 & 0.112 & 0.034 & - & L,X & 6,7 & 38\\ 
 33454 &            1999FJ27 &  2.3157 & 0.1483 & 0.0743 &    2.844 &  0.652 & 0.218 & 0.085 & - & C,CX & 6,7 & 38\\ 
 33528 &            Jinzeman &  2.3348 & 0.1664 & 0.0630 &    2.860 &  0.159 & 0.113 & 0.035 & - & X & 7 & 38\\ 
 33595 &            1999JC49 &  2.4731 & 0.1220 & 0.0316 &    3.972 &  0.106 & 0.123 & 0.022 & - & X & 7 & 38\\ 
 33597 &            1999JQ49 &  2.3882 & 0.1621 & 0.0228 &    2.173 &  0.435 & 0.236 & 0.103 & - & X & 7 & 38\\ 
 33901 &            2000KJ56 &  2.4693 & 0.1683 & 0.1756 &    4.695 &  0.105 & 0.184 & 0.044 & - & K,X & 6,7 & 38\\ 
 34120 &            2000PL28 &  2.3659 & 0.1810 & 0.1168 &    2.390 &  0.577 & 0.235 & 0.095 & - & C,X & 6,7 & 38\\ 
 34779 &        Chungchiyung &  2.2813 & 0.1311 & 0.1115 &    3.516 &  0.079 & 0.143 & 0.024 & - & X,Ds & 7,8 & 38\\ 
 35160 &              1993NY &  2.3140 & 0.1386 & 0.0370 &    3.435 &  0.062 & 0.111 & 0.033 & - & CX & 7 & 38\\ 
 36455 &             2000QZ6 &  2.4911 & 0.2493 & 0.2771 &    3.673 &  0.907 & 0.131 & 0.049 & - & X,C & 6,7 & 38\\ 
 36881 &           2000SX154 &  2.4147 & 0.1397 & 0.1793 &    4.381 &  0.163 & 0.278 & 0.038 & - & C,X & 6,7 & 38\\ 
 37723 &            1996TX28 &  2.4829 & 0.0796 & 0.0951 &    2.556 &  0.613 & 0.296 & 0.188 & - & Xt & 8 & 38\\ 
 37738 &            1996VM14 &  2.4410 & 0.1708 & 0.0174 &    2.161 &  0.441 & 0.287 & 0.115 & - & K,X & 6,7 & 38\\ 
 37982 &           1998HB132 &  2.4775 & 0.0199 & 0.0828 &    3.185 &  0.709 & 0.209 & 0.058 & - & C,X & 6,7 & 38\\ 
 38155 &            1999JJ69 &  2.2712 & 0.1843 & 0.0915 &    2.440 &  0.082 & 0.297 & 0.044 & - & X & 6 & 38\\ 
 38810 &            2000RP70 &  2.4631 & 0.1231 & 0.0373 &    4.509 &  0.207 & 0.115 & 0.038 & - & C,X & 6,7 & 38\\ 
 39038 &            2000UE80 &  2.3528 & 0.1366 & 0.0779 &    4.360 &  0.052 & 0.156 & 0.022 & - & C,X & 6,7 & 38\\ 
 39108 &            2000WG26 &  2.3293 & 0.1695 & 0.0702 &    3.077 &  0.102 & 0.107 & 0.027 & - & X,X & 6,7 & 38\\ 
 41817 &            2000WX40 &  2.2533 & 0.1305 & 0.1163 &    3.480 &  0.340 & 0.231 & 0.059 & - & Xt & 8 & 34\\ 
 42960 &           1999TJ139 &  2.4184 & 0.1339 & 0.0334 &    1.611 &  0.144 & 0.205 & 0.042 & - & S,X & 6,7 & 38\\ 
 43282 &           2000EB140 &  2.2392 & 0.1268 & 0.1027 &    2.867 &  0.067 & 0.163 & 0.015 & - & CX & 7 & 38\\ 
 43387 &            2000WF58 &  2.2990 & 0.1870 & 0.0867 &    2.520 &  0.389 & 0.254 & 0.099 & - & X & 7 & 38\\ 
 46603 &            1993FY41 &  2.4226 & 0.1257 & 0.0321 &    3.170 &  0.380 & 0.146 & 0.039 & - & X,CX & 6,7 & 34\\ 
 49701 &             1999VZ1 &  2.2108 & 0.1295 & 0.0308 &    1.682 &  0.160 & 0.207 & 0.043 & - & X,CX & 6,7 & 38\\ 
 49874 &           1999XW129 &  2.2405 & 0.1501 & 0.1206 &    2.467 &  0.301 & 0.201 & 0.050 & - & C,CX & 6,7 & 38\\ 
 49911 &           1999XT169 &  2.4184 & 0.2244 & 0.0788 &    4.216 &  0.293 & 0.208 & 0.030 & - & C,CX & 6,7 & 38\\ 
 50228 &           2000AD242 &  2.4582 & 0.1655 & 0.1725 &    3.640 &  0.290 & 0.193 & 0.037 & - & S,X & 6,7 & 34\\ 
 51440 &            2001FW24 &  2.2980 & 0.2377 & 0.1250 &    2.898 &  0.587 & 0.175 & 0.090 & - & X & 7 & 38\\ 
 52328 &            1992EK11 &  2.4026 & 0.1753 & 0.0878 &    1.865 &  0.289 & 0.292 & 0.062 & - & S,CX & 6,7 & 38\\ 
 52441 &             1994RS1 &  2.3990 & 0.2436 & 0.1702 &    2.287 &  0.058 & 0.281 & 0.045 & - & X & 7 & 38\\ 
 53625 &            2000CZ96 &  2.1523 & 0.0170 & 0.0862 &    2.138 &  0.581 & 0.293 & 0.120 & - & CX & 7 & 38\\ 
 53977 &            2000GM70 &  2.2461 & 0.1614 & 0.1099 &    3.156 &  0.076 & 0.173 & 0.013 & - & X,C,Ad & 6,7,8 & 38\\ 
 54103 &             2000HX6 &  2.4414 & 0.1772 & 0.0466 &    2.303 &  0.246 & 0.192 & 0.039 & - & CX & 7 & 38\\ 
 54355 &            2000KJ33 &  2.2760 & 0.1604 & 0.1173 &    4.496 &  0.054 & 0.139 & 0.009 & - & X & 7 & 38\\ 
 54419 &            2000LA20 &  2.4018 & 0.2234 & 0.2246 &    4.275 &  0.017 & 0.283 & 0.021 & - & X,X & 6,7 & 38\\ 
 55374 &           2001SE244 &  2.4551 & 0.0908 & 0.0267 &    2.667 &  0.718 & 0.108 & 0.077 & - & X & 7 & 38\\ 
 55467 &           2001TH173 &  2.2872 & 0.1905 & 0.0297 &    2.559 &  0.135 & 0.142 & 0.011 & - & S,X,S & 6,7,8 & 38\\ 
 55663 &             6247P-L &  2.2880 & 0.1119 & 0.0697 &    1.864 &  0.284 & 0.202 & 0.051 & - & X,C & 6,7 & 38\\ 
 56374 &            2000EM24 &  2.2262 & 0.1615 & 0.0880 &    1.591 &  0.277 & 0.211 & 0.056 & - & X & 7 & 38\\ 
 56408 &            2000FH22 &  2.1847 & 0.0624 & 0.0983 &    1.765 &  0.242 & 0.248 & 0.042 & - & C,X & 6,7 & 38\\ 
 56651 &            2000KH46 &  2.2388 & 0.1279 & 0.0940 &    1.595 &  0.504 & 0.159 & 0.065 & - & K,X,V & 6,7,8 & 38\\ 
 57434 &            2001SH46 &  2.3178 & 0.1978 & 0.0384 &    3.105 &  0.927 & 0.183 & 0.149 & - & S,X & 6,7 & 38\\ 
 58336 &              1994VP &  2.2722 & 0.1820 & 0.1140 &    2.352 &  0.226 & 0.153 & 0.089 & - & X & 7 & 38\\ 
 58598 &            1997TX11 &  2.2946 & 0.1499 & 0.0630 &    2.336 &  0.104 & 0.186 & 0.036 & - & X,CD,Kl & 6,7,8 & 38\\ 
 59024 &           1998SB106 &  2.3109 & 0.1720 & 0.0513 &    1.577 &  0.485 & 0.214 & 0.087 & - & C,X & 6,7 & 38\\ 
 59239 &             Alhazen &  2.2512 & 0.1888 & 0.1195 &    3.125 &  0.661 & 0.181 & 0.145 & - & X,C & 6,7 & 38\\ 
 59397 &            1999FT26 &  2.4372 & 0.1758 & 0.0460 &    2.855 &  0.192 & 0.198 & 0.021 & - & X,CX & 6,7 & 38\\ 
 60540 &            2000EZ61 &  2.2557 & 0.1522 & 0.0931 &    1.890 &  0.467 & 0.136 & 0.056 & - & C,X & 6,7 & 38\\ 
 60665 &            2000FL73 &  2.4540 & 0.1669 & 0.0421 &    1.982 &  0.356 & 0.284 & 0.068 & - & S,L,Xt & 6,7,8 & 38\\ 
 60916 &            2000JL37 &  2.2494 & 0.1309 & 0.1157 &    3.001 &  0.421 & 0.215 & 0.071 & - & C,X & 6,7 & 38\\ 
 63380 &            2001HE51 &  2.4598 & 0.0729 & 0.0919 &    1.938 &  0.560 & 0.225 & 0.133 & - & C,X & 6,7 & 38\\ 
 65745 &            1993TT31 &  2.2675 & 0.1545 & 0.0425 &    1.357 &  0.321 & 0.241 & 0.092 & - & Xt & 8 & 38\\ 
 67411 &            2000QJ26 &  2.2855 & 0.1261 & 0.1076 &    1.488 &  0.225 & 0.241 & 0.050 & - & CX & 7 & 38\\ 
 67597 &           2000SA141 &  2.2427 & 0.1365 & 0.0923 &    1.953 &  0.476 & 0.140 & 0.094 & - & C,X & 6,7 & 38\\ 
 67639 &           2000SE216 &  2.2508 & 0.1952 & 0.1092 &    1.772 &  0.404 & 0.170 & 0.047 & - & X & 7 & 38\\ 
 68550 &            2001XA54 &  2.3365 & 0.1876 & 0.0396 &    1.453 &  0.266 & 0.277 & 0.050 & - & X,X & 6,7 & 38\\ 
 68836 &           2002GU105 &  2.4477 & 0.1770 & 0.2158 &    4.398 &  0.145 & 0.110 & 0.027 & - & X & 7 & 38\\ 
 70125 &              1999NZ &  2.3502 & 0.1375 & 0.1778 &    3.632 &  0.084 & 0.102 & 0.013 & - & X,X & 6,7 & 38\\ 
 75374 &            1999XG84 &  2.4477 & 0.1434 & 0.1902 &    3.368 &  0.080 & 0.171 & 0.026 & - & Xt & 8 & 38\\ 
 75660 &            2000AG77 &  2.4408 & 0.1509 & 0.0760 &    3.722 &  0.207 & 0.127 & 0.041 & - & C,X & 6,7 & 38\\ 
 78647 &            2002TQ48 &  2.2443 & 0.1833 & 0.0929 &    2.576 &  0.525 & 0.184 & 0.050 & - & X & 7 & 38\\ 
 80650 &           2000AY246 &  2.2359 & 0.1652 & 0.0883 &    2.179 &  0.570 & 0.162 & 0.102 & - & C,CX & 6,7 & 38\\ 
 82676 &            2001PV23 &  2.3489 & 0.2670 & 0.1286 &    3.600 &  0.500 & 0.140 & 0.040 & - & X,CX,Kl & 6,7,8 & 71\\ 
 86797 &           2000GM108 &  2.2317 & 0.1584 & 0.0998 &    2.563 &  0.061 & 0.269 & 0.033 & - & X,V & 7,8 & 38\\ 
 86799 &           2000GH112 &  2.2346 & 0.1529 & 0.0802 &    2.069 &  0.320 & 0.217 & 0.043 & - & X & 7 & 38\\ 
 88629 &            2001RQ34 &  2.2462 & 0.1370 & 0.0672 &    2.256 &  0.317 & 0.166 & 0.094 & - & X,D & 6,7 & 38\\ 
 90084 &           2002VC116 &  2.3493 & 0.2362 & 0.1351 &    3.446 &  0.061 & 0.196 & 0.025 & - & C,X & 6,7 & 38\\ 
 90102 &            2002XQ21 &  2.2741 & 0.2260 & 0.0672 &    3.170 &  0.159 & 0.133 & 0.045 & - & X & 7 & 38\\ 
 90115 &            2002XJ54 &  2.2327 & 0.0957 & 0.0777 &    1.641 &  0.215 & 0.238 & 0.071 & - & C,CX & 6,7 & 38\\ 
 91242 &            1999CX32 &  2.3018 & 0.1701 & 0.1472 &    2.529 &  0.099 & 0.230 & 0.047 & - & C,X & 6,7 & 38\\ 
 91309 &            1999FB55 &  2.3557 & 0.1740 & 0.1729 &    3.313 &  0.605 & 0.161 & 0.059 & - & C,X & 6,7 & 38\\ 
 92324 &            2000GE49 &  2.2156 & 0.1344 & 0.1056 &    1.526 &  0.172 & 0.209 & 0.034 & - & X & 7 & 38\\ 
 92537 &            2000OS16 &  2.2419 & 0.1539 & 0.1268 &    2.275 &  0.324 & 0.215 & 0.035 & - & X & 7 & 38\\ 
 94688 &            2001XL27 &  2.3978 & 0.1563 & 0.2051 &    2.864 &  0.413 & 0.236 & 0.080 & - & Xt & 8 & 38\\ 
 96288 &             1996GD6 &  2.2389 & 0.1340 & 0.0940 &    2.052 &  0.438 & 0.116 & 0.040 & - & C,CX & 6,7 & 38\\ 
 96349 &             1997US7 &  2.2485 & 0.1238 & 0.0169 &    2.271 &  0.318 & 0.124 & 0.032 & - & X,CX & 6,7 & 38\\ 
 96651 &            1999GT62 &  2.3549 & 0.1458 & 0.1379 &    2.434 &  0.143 & 0.156 & 0.035 & - & X,C,C & 6,7,8 & 38\\ 
 98089 &            2000RO72 &  2.2882 & 0.1698 & 0.0753 &    2.023 &  0.342 & 0.188 & 0.064 & - & X & 7 & 38\\ 
 98123 &            2000SG15 &  2.2272 & 0.1222 & 0.0820 &    1.922 &  0.243 & 0.158 & 0.058 & - & C,CX & 6,7 & 38\\ 
 98352 &           2000SX327 &  2.3422 & 0.1509 & 0.0780 &    2.647 &  0.637 & 0.191 & 0.076 & - & X,C & 6,7 & 38\\ 
 98383 &            2000TL39 &  2.2361 & 0.1475 & 0.1178 &    2.420 &  0.691 & 0.190 & 0.092 & - & X,C & 6,7 & 38\\ 
 99650 &             2002HF2 &  2.2882 & 0.0366 & 0.0552 &    1.780 &  0.298 & 0.202 & 0.061 & - & X,Ad & 7,8 & 38\\ 
103600 &            2000CM16 &  2.2358 & 0.1351 & 0.1051 &    3.099 &  0.691 & 0.106 & 0.043 & - & S,X & 6,7 & 38\\ 
105983 &           2000SE269 &  2.2713 & 0.1634 & 0.1039 &    2.170 &  0.242 & 0.164 & 0.049 & - & S,X & 6,7 & 38\\ 
109038 &            2001QC13 &  2.4699 & 0.1665 & 0.0502 &    2.081 &  0.355 & 0.195 & 0.056 & - & X,C,Ad & 6,7,8 & 38\\ 
111483 &            2001YZ38 &  2.2276 & 0.1354 & 0.0960 &    1.640 &  0.382 & 0.165 & 0.089 & - & S,X & 6,7 & 38\\ 
112019 &           2002GL168 &  2.2531 & 0.0755 & 0.0908 &    1.947 &  0.211 & 0.107 & 0.031 & - & X & 7 & 38\\ 
120005 &            2002YB31 &  2.3768 & 0.1792 & 0.0440 &    1.478 &  0.272 & 0.203 & 0.062 & - & X,XL & 6,7 & 38\\ 
121951 &            2000ES45 &  2.3505 & 0.2060 & 0.1770 &    2.907 &  0.858 & 0.120 & 0.064 & - & X & 7 & 38\\ 
129428 &             4164T-3 &  2.2734 & 0.1144 & 0.0681 &    2.747 &  0.599 & 0.123 & 0.092 & - & X,CX & 6,7 & 38\\ 
130744 &           2000SH259 &  2.2835 & 0.0964 & 0.0447 &    1.919 &  0.456 & 0.132 & 0.061 & - & CX & 7 & 38\\ 
132366 &            2002GF68 &  2.4026 & 0.1823 & 0.1937 &    1.989 &  0.536 & 0.234 & 0.191 & - & K,X & 6,7 & 38\\ 
136797 &             1997CF1 &  2.2131 & 0.1994 & 0.1346 &    1.309 &  0.386 & 0.179 & 0.101 & - & X & 7 & 38\\ 
137818 &             2000AR3 &  2.4150 & 0.1685 & 0.2391 &    2.948 &  0.047 & 0.117 & 0.015 & - & X,D & 6,7 & 38\\ 
145882 &            1999TU32 &  2.2207 & 0.1623 & 0.0446 &    1.342 &  0.100 & 0.270 & 0.071 & - & X,C & 6,7 & 38\\ 
148158 &           1999XB153 &  2.2357 & 0.1478 & 0.0907 &    2.016 &  0.379 & 0.158 & 0.045 & - & C,CX & 6,7 & 38\\ 
149614 &            2004EP11 &  2.2300 & 0.1295 & 0.0884 &    2.017 &  0.256 & 0.109 & 0.044 & - & CX & 7 & 38\\ 
155998 &            2001RQ21 &  2.2387 & 0.1016 & 0.0211 &    1.860 &  0.111 & 0.128 & 0.016 & - & X & 7 & 38\\ 
158611 &            2003AY35 &  2.2622 & 0.1694 & 0.0720 &    1.768 &  0.237 & 0.118 & 0.045 & - & X,C & 6,7 & 38\\ 
161242 &            2003BK13 &  2.3695 & 0.1559 & 0.1780 &    2.456 &  0.088 & 0.106 & 0.019 & - & X,C & 6,7 & 38\\ 
162528 &           2000QT134 &  2.2325 & 0.1317 & 0.1109 &    1.912 &  0.409 & 0.192 & 0.126 & - & X,C & 6,7 & 38\\ 
166079 &           2002CW102 &  2.3256 & 0.2166 & 0.0512 &    1.888 &  0.507 & 0.180 & 0.108 & - & X,C & 6,7 & 38\\ 
167102 &            2003SB47 &  2.4350 & 0.1960 & 0.1293 &    1.880 &  0.616 & 0.165 & 0.098 & - & S,CX & 6,7 & 38\\ 
188198 &            2002RX59 &  2.2528 & 0.1600 & 0.0872 &    1.714 &  0.241 & 0.218 & 0.126 & - & C,X,S & 6,7,8 & 38\\ 
193024 &            2000ET60 &  2.2387 & 0.1800 & 0.0998 &    1.489 &  0.335 & 0.182 & 0.104 & - & X,C & 6,7 & 38\\ 
194798 &            2001YW97 &  2.3405 & 0.1800 & 0.0580 &    2.296 &  0.180 & 0.160 & 0.026 & - & X & 7 & 38\\ 
215473 &            2002RZ50 &  2.2390 & 0.1574 & 0.1151 &    2.103 &  0.472 & 0.159 & 0.071 & - & K,X & 6,7 & 38\\ 
222425 &              2001OV &  2.4498 & 0.1977 & 0.1232 &    1.532 &  0.426 & 0.189 & 0.104 & - & X & 7 & 38\\ 
227423 &            2005VC42 &  2.4477 & 0.1247 & 0.1882 &    2.085 &  0.270 & 0.162 & 0.029 & - & CX & 7 & 38\\ 
234143 &            2000EY44 &  2.2249 & 0.1424 & 0.0985 &    1.730 &  0.353 & 0.135 & 0.040 & - & C,X & 6,7 & 38\\ 
235005 &            2003CW17 &  2.2143 & 0.1567 & 0.0951 &    1.210 &  0.200 & 0.121 & 0.042 & - & CX & 7 & 34\\ 
235039 &            2003FN40 &  2.3352 & 0.1436 & 0.0690 &    1.601 &  0.309 & 0.131 & 0.049 & - & X & 7 & 38\\ 
240308 &           2003FP100 &  2.2898 & 0.1288 & 0.0874 &    1.316 &  0.296 & 0.147 & 0.053 & - & K,X & 6,7 & 38\\ 
240319 &            2003HU37 &  2.3019 & 0.1711 & 0.1901 &    2.687 &  0.148 & 0.117 & 0.011 & - & Xt & 8 & 38\\ 
241921 &           2002AG186 &  2.4654 & 0.1489 & 0.1727 &    1.811 &  0.422 & 0.135 & 0.035 & - & C,CX & 6,7 & 38\\ 
242248 &           2003ST233 &  2.3996 & 0.2654 & 0.1682 &    1.856 &  0.413 & 0.117 & 0.018 & - & X,L,S & 6,7,8 & 38\\ 
243780 &            2000SC20 &  2.3671 & 0.1808 & 0.1366 &    2.072 &  0.436 & 0.124 & 0.055 & - & X & 7 & 38\\ 
253490 &            2003SJ91 &  2.2921 & 0.1289 & 0.0969 &    1.453 &  0.248 & 0.121 & 0.061 & - & X,X & 6,7 & 38\\ 
276825 &           2004PM101 &  2.2125 & 0.1955 & 0.1549 &    1.400 &  0.200 & 0.120 & 0.040 & - & C,X & 6,7 & 71\\ 
423025 &            2003TB21 &  2.2956 & 0.1517 & 0.1054 &    1.293 &  0.253 & 0.116 & 0.029 & - & C,CX & 6,7 & 38\\ 

\hline
\hline
\end{longtable}

\newpage

\begin{longtable}{l c c c}
  \caption{\label{T:Xdensities}
Known densities and their uncertainties of asteroids of the X-complex and with 0.1 $< p_V<$ 0.3. Those asteroids with unrealistic densities, marked with a cross in the Tab.~1 of \cite{Carry2012P&SS...73...98C} are not included. The column Ref. contains the reference to the publication, as follow: a = \cite{Hanus2017A&A...601A.114H}, b = \cite{Avdellidou2018MNRAS.475.3419A}, c = \cite{Sierks2011Sci...334..487S}, d = \cite{Carry2012P&SS...73...98C}. }\\
\hline
Asteroid & Density & $1\sigma$ & Ref. \\
& kg~m$^{-3}$ & kg~m$^{-3}$ & \\
\hline
  16 Psyche    & 3,700 & 600 & a \\       
  15  Eunomia  & 3,500 & 400 &  b \\   
  21   Lutetia      & 3,400 & 300 & c \\	    
  22    Kalliope   & 3,700 & 400 & a \\	    
  69    Hesperia & 4,400 & 1,000 & d \\	    
  97     Klotho    & 4,200 & 600 & d \\	    
129   Antigone  & 2,500 & 900 & a \\	    
135        Hertha & 4,500 & 700 & a \\	    
216   Kleopatra & 5,000 & 700 & a \\	    
516   Amherstia & 8,000 & 7,600 & d \\	    
665        Sabine & 9,100 & 5,200 & d \\
758    Mancunia & 2,700 & 300 & d \\ 
\hline    
\end{longtable}

\begin{longtable}{rr|ccccccccccc}
\caption{\label{T:planetesimals}
Original asteroids i.e. the surviving planetesimals. The first group in the table is from \cite{Delbo2017Sci...357.1026M}. The second from this work.
For references see Tab~\ref{T:AthorMembers}.}\\
\hline\hline
Number & Designation & $a$ & $e$ & sin($i$) & $D$ & $\sigma_D$ & $p_v$ & $\sigma_{p_V}$ & $P$ & Taxonomic & Class & $D, p_V$\\ 
 &  & (au) & & & (km) & (km) & &  & (hour) & class  & reference & reference\\ 
\hline
     4 &               Vesta &  2.3615 & 0.0988 & 0.1113 &  521.740 &  7.500 & 0.342 & 0.013 &   5.34 & V,V,V & 1,2,4 & 37\\ 
     8 &               Flora &  2.2014 & 0.1449 & 0.0971 &  147.490 &  1.030 & 0.230 & 0.040 &  12.86 & Sw,S & 1,4 & 39\\ 
    18 &           Melpomene &  2.2958 & 0.1802 & 0.1701 &  139.590 &  2.450 & 0.230 & 0.030 &  11.57 & S,S,S & 1,2,4 & 39\\ 
    51 &             Nemausa &  2.3657 & 0.1140 & 0.1740 &  138.160 &  0.970 & 0.100 & 0.030 &   7.78 & Cgh,Ch,CU & 1,2,4 & 39\\ 
   654 &             Zelinda &  2.2970 & 0.2130 & 0.3165 &  116.300 &  2.380 & 0.050 & 0.010 &  31.74 & Ch,C & 2,4 & 39\\ 
    12 &            Victoria &  2.3343 & 0.1751 & 0.1624 &  115.090 &  1.200 & 0.160 & 0.030 &   8.66 & L,S,D & 2,4,5 & 39\\ 
    27 &             Euterpe &  2.3470 & 0.1865 & 0.0123 &  114.100 &  4.480 & 0.220 & 0.030 &  10.41 & S,S,S,L & 1,2,4,5 & 39\\ 
    40 &            Harmonia &  2.2673 & 0.0216 & 0.0654 &  111.250 &  0.390 & 0.220 & 0.050 &   8.91 & S,S,S & 1,2,4 & 39\\ 
   345 &           Tercidina &  2.3253 & 0.0987 & 0.1799 &   99.000 & 11.469 & 0.059 & 0.012 &  12.37 & Ch,Ch,C & 1,2,4 & 38\\ 
   326 &              Tamara &  2.3176 & 0.2034 & 0.3943 &   89.420 &  1.510 & 0.040 & 0.002 &  14.44 & C & 4 & 37\\ 
    72 &             Feronia &  2.2662 & 0.0741 & 0.1037 &   78.800 &  2.000 & 0.080 & 0.010 &   8.10 & STD,TDG & 3,4 & 39\\ 
    80 &              Sappho &  2.2960 & 0.1508 & 0.1620 &   74.250 &  3.000 & 0.210 & 0.010 &  14.03 & S,S & 2,4 & 39\\ 
   336 &           Lacadiera &  2.2518 & 0.0882 & 0.1071 &   69.000 &  3.364 & 0.046 & 0.005 &  13.70 & Xk,D & 2,4 & 38\\ 
   207 &               Hedda &  2.2840 & 0.0656 & 0.0589 &   57.880 &  0.150 & 0.060 & 0.010 &  30.10 & Ch,C,Ch & 2,4,5 & 39\\ 
   261 &              Prymno &  2.3315 & 0.1311 & 0.0512 &   50.010 &  0.520 & 0.140 & 0.030 &   8.00 & X,B & 2,4 & 39\\ 
   136 &             Austria &  2.2868 & 0.0183 & 0.1732 &   36.890 &  0.520 & 0.220 & 0.050 &  11.50 & Xe,M & 2,4 & 39\\ 
   376 &           Geometria &  2.2886 & 0.1712 & 0.1064 &   35.470 &  0.050 & 0.320 & 0.050 &   7.74 & Sl,S,S,S & 2,4,6,7 & 39\\ 
   298 &          Baptistina &  2.2639 & 0.1476 & 0.1038 &   20.530 &  0.270 & 0.170 & 0.005 &  16.23 & Xc & 5 & 37\\ 
\hline
\hline
    21 &             Lutetia &  2.4353 & 0.1292 & 0.0374 &  108.380 &  1.280 & 0.181 & 0.005 &   8.17 & Xc,Xk,M,X & 1,2,4,5 & 37\\ 
   131 &                Vala &  2.4316 & 0.0940 & 0.0721 &   31.340 &  0.300 & 0.170 & 0.040 &   5.18 & K,Xc,CX,SU & 1,2,3,4 & 39\\ 
   135 &              Hertha &  2.4285 & 0.1741 & 0.0466 &   71.040 &  2.650 & 0.180 & 0.030 &   8.40 & Xk,M & 2,4 & 39\\ 
   261 &              Prymno &  2.3315 & 0.1311 & 0.0512 &   50.010 &  0.520 & 0.140 & 0.030 &   8.00 & X,B & 2,4 & 39\\ 
   337 &              Devosa &  2.3830 & 0.1569 & 0.1391 &   66.630 &  0.980 & 0.127 & 0.005 &   4.65 & Xk,X,X & 1,2,4 & 37\\ 
   435 &                Ella &  2.4495 & 0.1212 & 0.0287 &   37.040 &  0.500 & 0.106 & 0.003 &   4.62 & DCX & 4 & 37\\ 
   572 &             Rebekka &  2.4007 & 0.1358 & 0.1854 &   26.190 &  0.370 & 0.111 & 0.004 &   5.65 & C,XDC,C,C & 2,4,6,7 & 37\\ 
 13977 &              Frisch &  2.4715 & 0.0797 & 0.2768 &    9.260 &  0.110 & 0.160 & 0.020 &   5.00 & X,X & 6,7 & 39\\ 

\hline
\hline
\end{longtable}

\end{small}

\end{appendix}

\end{document}